\documentclass[preprint2]{aastex}

\newcommand{\myemail}{sdzib@mpifr-bonn.mpg.de}

\slugcomment{Second Draft}

\shorttitle{A VLA survey of Taurus-Auriga}
\shortauthors{Dzib et al.}

\usepackage{epsfig}
\usepackage{longtable}
\usepackage{amssymb}
\usepackage{lscape}

\begin{document}

\newpage

\title{The Gould Belt Very Large Array Survey IV:\\
The Taurus-Auriga complex}

\author{Sergio A. Dzib\altaffilmark{1}, 
Laurent Loinard\altaffilmark{2}, 
Luis F.\ Rodr\'{\i}guez\altaffilmark{2,3},
Amy J.\ Mioduszewski\altaffilmark{4},  
Gisela N.\ Ortiz-Le\'on\altaffilmark{2}, 
Marina A. Kounkel\altaffilmark{5},
Gerardo Pech\altaffilmark{2}, 
Juana L.\ Rivera\altaffilmark{2}, 
Rosa M.\ Torres\altaffilmark{6}, 
Andrew F.\ Boden\altaffilmark{7}, 
Lee Hartmann\altaffilmark{5}, 
Neal J.\, Evans II\altaffilmark{8}, 
Cesar Brice\~no\altaffilmark{9} and
John Tobin\altaffilmark{10}
}

\altaffiltext{1}{Max Planck Institut f\"ur Radioastronomie, Auf dem H\"ugel 69, 53121 Bonn, Germany (\myemail)}

\altaffiltext{2}{Centro de Radioastronom\'{\i}a y Astrof\'{\i}sica, Universidad
Nacional Aut\'onoma de M\'exico\\ Apartado Postal 3-72, 58090,
Morelia, Michoac\'an, Mexico }

\altaffiltext{3}{King Abdulaziz University, P.O. Box 80203, Jeddah 21589, Saudi Arabia}

\altaffiltext{4}{National Radio Astronomy Observatory, Domenici Science Operations Center,\\
1003 Lopezville Road, Socorro, NM 87801, USA}

\altaffiltext{5}{Department of Astronomy, University of Michigan, 500 Church Street, Ann Arbor, MI 48105, USA}

\altaffiltext{6}{Instituto de Astronom\'{\i}a y Meteorolog\'{\i}a, Universidad de Guadalajara, 
Avenida Vallarta No. 2602, Col. Arcos Vallarta, CP 44130, Guadalajara, Jalisco, M\'exico}

\altaffiltext{7}{Division of Physics, Math, and Astronomy, California Institute of Technology, 1200 E California Blvd., 
Pasadena, CA 91125, USA}

\altaffiltext{8}{Department of Astronomy, The University of Texas at Austin, 1 University Station, C1400, Austin, 
TX 78712, USA}

\altaffiltext{9}{Cerro Tololo Interamerican Observatory, Casilla 603, La Serena, Chile.}

\altaffiltext{10}{Leiden Observatory, Leiden University, P.O. Box 9513, 2300 RA Leiden, The Netherlands }


\begin{abstract}
We present a multi-epoch radio study of the Taurus-Auriga star-forming complex
made with the Karl G. Jansky Very Large Array at frequencies of 4.5 GHz and 7.5 GHz.
We detect a total of 610 sources, 59 of which are related to young stellar
objects and 18 to field stars. The properties of 56\% of the young stars are compatible with
non-thermal radio emission. We also show that the radio emission of more
evolved young stellar objects tends to be more non-thermal in origin and, 
in general, that their radio properties are compatible with those found in other
star forming regions. By comparing our results with previously reported 
X-ray observations, we notice that young stellar objects in Taurus-Auriga
follow a G\"{u}del-Benz relation with $\kappa$=0.03, as we previously
suggested for other regions of star formation. In general, young stellar objects 
in Taurus-Auriga and in all the previous studied regions seem to follow this relation
with a dispersion of $\sim1$ dex. 
Finally, we propose that most of the remaining sources are related with extragalactic
objects but provide a list of 46 unidentified radio sources whose radio properties are 
compatible with a YSO nature. 
\end{abstract}


\section{Introduction}

Taurus-Auriga is one of the best studied star-forming regions. 
Indeed, many of the pioneering studies of star formation were 
based on observation of the Young Stellar Objects (YSOs)
and clouds that compose this complex (see Kenyon
et al. 2008, for a recent review). The Taurus-Auriga region is
particularly appropriate to study invididual low-mass YSOs because
star-formation there occurs in a quiescent environment with low 
stellar density. This is unlike the situation for stars forming in dense
clusters exposed to massive stars such {  as in the Orion Nebula region}.
In addition, its relative proximity (140 pc with a depth of up 
to 30 pc, see Loinard et al.\ 2005, 2007a; Torres et al.\ 2007, 
2009, 2012) makes this complex an ideal candidate for
studies of individual YSOs.

\begin{figure*}[b!]
\begin{center}
\includegraphics[width=0.70\textwidth,angle=-90]{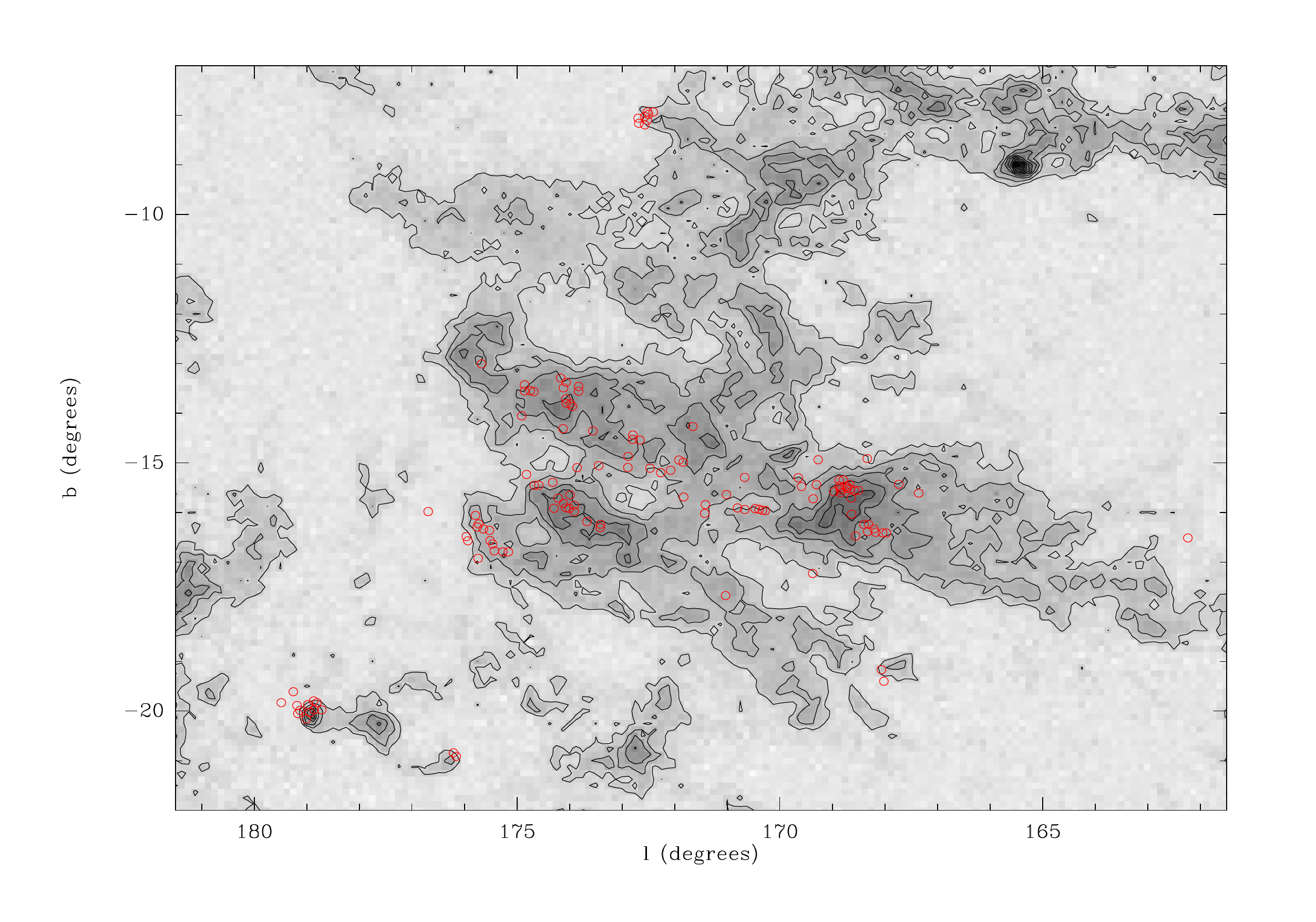}
\end{center}
\caption{Background: CO extinction map (from Dobashi et al.~2005). 
Red circles are the VLA observed fields. The diameter of each circle is 10$'$,
i.e. slightly larger than the primary beam used at 7.5 GHz, see text.}
\label{fig:ma+fields}
\end{figure*}

Observations in various bands of the electromagnetic spectrum 
have been obtained for the entire complex (e.g., 
X-rays by G\"{u}del et al.~2007; 
optical by Brice\~no et al. 1993, 1999;
infrared by {  Padgett et al.~2007}; 
near-infrared by Duch{\^e}ne et al.~2004, and
sub-millimeter by Andrews \& Williams~2005). In contrast, 
however, there are very few radio surveys. This is mainly due to the 
large solid angle covered by the Taurus-Auriga complex (more  
than 100 degrees$^2$; e.g., Figure \ref{fig:ma+fields})
compared with the typical field of view of radio interferometers 
($\sim$0.1 degrees$^2$). In general, radio observations
have been devoted to some specific sources of interest.
An extensive survey was performed by O'Neal et al.
(1990) who pointed 99 fields with the VLA at a frequency
of 5 GHz to reach a sensitivity of 0.7 mJy. These
fields contained a total of 119 YSO candidates, but only nine 
were detected. Here we present a study
with the same approach but with one order of magnitude
better sensitivity than those obtained by O'Neal
et al.~(1990). Additionally, we simultaneously observe at two frequencies
and in three different epochs to characterize the spectral
index and the variability of the sources.

This research is part of a recently initiated radio survey
of YSOs in nearby ($<500$ pc) star forming regions (see 
Dzib et al.~2013a; Kounkel et al.~2014; Ortiz-Le\'on et 
al.~2014 submitted;  and Pech et al.~2014 in preparation, 
for recent results) to characterize their radio emission. Particularly, 
we are interested in distinguishing between thermal and 
non-thermal emission of YSOs by measuring the 
spectral index $\alpha$ (defined such that the flux density
depends on frequency as S$_{\nu}$ $\propto$ $\nu^{\alpha}$)
and variability of the flux density. These data will also be used to
compare the radio emission properties with the emission at
other wavelengths.

\section{Observations}

The observations were obtained with the Karl G. Jansky Very Large Array (VLA)
of the National Radio Astronomy Observatory (NRAO) in its B and BnA configuration. 
Two frequency sub-bands, each 1 GHz wide, and centered at 4.5 and 7.5 GHz, 
respectively, were recorded simultaneously.  The observations were obtained on three 
different time periods  (February 25/26/28 to March 06; April 12/17/20/25, and  
April 30 to May 1/5/14/22 on 2011) typically separated from one another by a month,
see Table \ref{tab:obs} for details. This dual frequency, multi-epoch
strategy was chosen to enable the characterization of the spectral index and
variability of the detected sources, and to help in the identification of the
emission mechanisms (thermal vs. non-thermal). 


For our study, we observed 127 different target fields distributed
across the cloud complex (Figure \ref{fig:ma+fields}). The fields
were chosen to cover previously known YSOs. In 33 of those fields,
 we could observe more than one {  YSO} target, while in the remaining 94 
fields, only one YSO was targeted. In most cases, the infrared 
evolutionary class (i.e., Class I, II or III) or T Tauri evolutionary status 
(classical or weak line) of the targetted sources was known from the
literature. Due to the number of observed fields we divided them in 
five different blocks, {named from A to E}, and each observation session {  ran} for two hours.
We obtained a total of 15 observation sessions for the Taurus-Aurigae
region, and they are summarized in Table \ref{tab:obs}.

Because of the large total area covered by the observations, three
different phase calibrators were used: J0449+1121, J0403+2600 and 
J0443+3441, depending on the position of the target fields. The flux
calibrator, for all the observations, was the quasar 3C~147 that was
 also used as the bandpass calibrator.

Each observing session was organized as follows. The standard flux calibrator 
3C~147 was first observed for $\sim$10 minutes. We subsequently spent one 
minute on the phase calibrator followed by a series of three target
pointings, spending three minutes on each, closing the cycle with a new 
observation of one minute on the phase calibrator. This phase calibrator/target sequence
was repeated until all target fields were observed. Thus, three minutes were
spent on each target field for each epoch. The data were edited, calibrated, and 
imaged in a standard fashion using the Common Astronomy Software Applications 
package (CASA). {The data calibration was the same as that performed for the Ophiuchus
observations reported in Dzib et al.\ (2013a) and is not repeated here. 
When imaging, the minimum level of the primary beam used was at a response of
20\%. The final images covered circular areas of 8.8 and 14.3 arcminutes in
diameter, for the 7.5 and 4.5 GHz sub-bands, respectively, and were corrected
for the effects of the position-dependent primary beam response.
The noise levels reached for each individual observation was about 
$\sim40~\mu$Jy and $\sim30~\mu$Jy, at 4.5 GHz and 7.5 GHz, respectively (see
Table \ref{tab:obs} for the individual values at each epoch). 

In two epochs we detected problems. First, for the third epoch of Block A 
we observed {  systematically} lower {  flux} densities for the detected
sources, compared to the first two epochs, thus we discard this epoch for our main
analysis. Still, we used this epoch to search any extra detections. Second, at
the beginning of the observation of the second epoch of Block C, some scans were 
missed. Despite these problems, our analysis is not affected.

{  The visibilities of the three, or two, observations obtained for each field
were concatenated to produce a new image with a lower noise level }
 (of about $\sim25~\mu$Jy at 4.5 GHz and $\sim18~\mu$Jy at 7.5 GHz).
The angular resolution of  $\sim~1''$ (see the synthesized beam sizes in Table 
\ref{tab:obs}) allows an uncertainty in position of $\sim~$0\rlap{$''$}.\,1 or better. 
To test for circular polarization we produced images 
of the $V$ Stokes parameter in the inner quarter (in area) of the primary beam at each frequency.
At larger distances from the field center, polarization measurements become unreliable as beam squint 
(the separation of the $R$ and $L$ beams on the sky) can create artificial circular
polarization signals.
 }
 

\begin{deluxetable}{cccccccc}
\tabletypesize{\scriptsize}
\tablewidth{0pt}
\tablecolumns{8}
\tablecaption{ Taurus-Auriga VLA observations. \label{tab:obs}}
\tablehead{     Block &    Epoch\tablenotemark{a}  &   Date  & VLA & \multicolumn{2}{c}{Synthesized beam}  & \multicolumn{2}{c}{rms noise} \\
\colhead{}&\colhead{}&\colhead{2011} & Config. &
\multicolumn{2}{c}{($\theta_{\rm maj}['']\times\theta_{\rm min}[''];$ P.A.[$^\circ$])} & \multicolumn{2}{c}{$\mu$Jy beam$^{-1}$}\\
\colhead{} & \colhead{} &\colhead{(dd.mm)}& &\colhead{4.5 GHz} & \colhead{7.5 GHz} & \colhead{4.5 GHz} & \colhead{7.5 GHz}\\
}\startdata
A & 1 & 25.02 & B & $1.17\times 1.09;\, 101.7$ & $0.72\times0.67;\, 108.4$ & 35 & 29 \\
  & 2 & 17.04 & B & $1.15\times 1.08;\, 104.3$ & $0.69\times0.65;\, 112.2$ & 39 & 32 \\
  & 3 & 14.05 & B$\rightarrow$A& $2.69\times 0.36;\, 58.8$ & $1.35\times 0.24;\, 59.7$ & 50 & 36 \\  
  & {\it C}\,\tablenotemark{b} &       & & $1.17\times 1.10;\,-76.8$ & $0.72\times 0.67;\,-72.0$ & 30 & 22 \\
B & 1 & 25.02 & B & $1.62\times 1.06;\, \,84.6$ & $1.02\times 0.64;\, \,84.2$ & 34 & 27 \\
  & 2 & 12.04 & B & $1.89\times 1.09;\,\,89.3$ & $1.53\times 0.63;\, \,85.2$ & 52 & 40 \\
  & 3 & 01.05 & B & $1.30\times 0.97;\, -20.6$ & $0.69\times 0.65;\, -49.7$ & 42 & 30 \\  
  & {\it C }&       &   & $1.37\times 1.11;\,\,90.2$ & $0.86\times 0.65;\, \,86.7$ & 26 & 19 \\
C & 1 & 26.02 & B & $1.10\times 1.02;\, -21.9$ & $0.68\times 0.63;\, -19.0$ & 35 & 27 \\
  & 2 & 20.04 & B & $1.66\times 1.02;\,\, 84.5$ & $1.03\times 0.63;\,\, 85.6$ & 38 & 30 \\
  & 3 & 05.05 & B & $1.66\times 1.10;\,\, 81.5$ & $1.02\times 0.65;\,\, 86.0$ & 39 & 30 \\  
  & {\it C} &       &   & $1.30\times 1.07;\,\, 89.3$ & $0.80\times 0.66;\,\, 90.4$ & 22 & 18 \\
D & 1 & 28.02 & B & $1.12\times 1.05;\,\, 44.7$ & $0.65\times 0.62;\,\, 42.1$ & 34 & 30 \\
  & 2 & 25.04 & B & $1.11\times 1.01; -24.2$ & $0.68\times 0.63;\, -22.7$ & 35 & 25 \\
  & 3 & 30.04 & B & $1.50\times 1.14;\,\,69.6$ & $0.89\times 0.70;\,\,74.8$ & 45 & 36 \\  
  & {\it C} &       &   & $1.13\times 1.09;\,\,56.5$ & $0.68\times 0.66;\,\,60.5$ & 21 & 17 \\
E & 1 & 06.03 & B & $1.38\times 1.07; -84.2$ & $0.85\times 0.65; -84.3$ & 32 & 26 \\
  & 2 & 25.04 & B & $1.11\times 1.02; -42.0$ & $0.66\times 0.60; -35.7$ & 39 & 32 \\  
  & 3 & 22.05 & B$\rightarrow$A&$1.68\times 0.36; -69.9$ & $1.06\times 0.22; -69.8$ & 40 & 31 \\
  & {\it C} &       &   & $1.32\times 0.53;\,  -71.2$ & $0.82\times 0.33;-70.0$ & 24 & 19 \\
\enddata
\tablenotetext{a}{{\it C } indicates parameters measured in the images after combining the epochs.}
\tablenotetext{b}{Only epoch 1 and 2 were concatenated due to {  systematically} lower flux densities values
on epoch 3. See also text.}
\end{deluxetable}

\section{Results}

In the observed area, there is a total of 196 known YSOs. 
{  The first step was the identification of radio sources in the observed
fields. We follow the procedure and criteria presented by Dzib et al.~(2013a)
who consider a detection as firm if the sources have a flux larger than 
4$\times\sigma_{\rm noise}$ level and there is a counterpart know at other 
wavelength, else we require a 5$\times\sigma_{\rm noise}$ level}.
The {  identification} was done using the images corresponding to the concatenation
of the observed epochs, which provides the highest sensitivity. 
From this, a total of 609 sources were detected (see Table
\ref{tab:sources}).  From these sources 215 were only detected in the 4.5 GHz
sub-band, while six were only detected in the 7.5 GHz sub-band. The 
remaining 388 sources were detected in both {  sub-bands}. 
We also used the third epoch of Block A to search for any additional detection.
One source, related to the YSO Coku HP Tau G3, was detected at both sub-bands 
in this epoch, but not detected in the concatenated map of the two first epochs
of Block A. Consequently, the total {  number of} detected sources in this survey is 610.
To reflect the fact that these sources were found
as part of the {\it Gould's Belt Very Large Array Survey}, a source
with coordinates {\it hhmmss.ss$+$ddmmss.s} will be named GBS-VLA
J{\it hhmmss.ss$+$ddmmss.s}.

The fluxes of each source at 4.5 and 7.5 GHz are given in columns 2 and
4 of Table \ref{tab:sources}. Two main sources of uncertainties on the
fluxes are included {  and given independently in Table \ref{tab:sources}}: 
(i) the error that results from the statistical
noise in the images and (ii) a systematic uncertainty of 5\% resulting
from possible errors in the absolute flux calibration {  (e.g., 
Perley \& Buttler 2013)}. We additionally
take into account errors produced by pointing errors of the primary
beam (following Dzib et al.~2014), whose effects are more noticeable
for sources detected at the edges of the primary beam. These errors
were added in quadrature to the statistical errors. An estimation of
the radio spectral index of each source (given in column 6 of Table
\ref{tab:sources}) was obtained from the fluxes measured in each
sub-band (at 4.5 and 7.5 GHz). To calculate the errors on the spectral
indices, {  also given in column 6 of Table \ref{tab:sources}}, the sources
of errors on the flux at each frequency were added in quadrature and 
the final error was obtained using standard error propagation theory.

Once the sources were identified in the concatenated images, we searched
for them in the images obtained from the individual epochs. An estimate 
of the level of variability of the sources was obtained by comparing the fluxes 
measured at the three epochs. Specifically, we calculated, for each source and at 
each frequency, the difference between the highest and lowest measured fluxes, 
and normalized by the maximum flux. The resulting values, {  expressed as 
percentages are given} in columns 3 and 5 of Table \ref{tab:sources}.
{  To calculate the errors on the variability, also given in columns 3 and 
5 of Table \ref{tab:sources}, the sources of errors on the
flux at each epoch were added in quadrature and the final error was
obtained using standard error propagation theory.} We will define
as a highly variable sources those that change their flux density by at least a factor
of two between the maximum and minimum measured fluxes (i.e., a variability larger
than 50\%) and that the result is significant above a three sigma level. Circular
polarization was confidently detected in 11 radio sources (Table \ref{tab:polz}).

Having identified the radio sources in the region mapped, and having derived their
variability, polarization, and spectral index, our next step was to try 
to determine which type of object they are associated with. In our specific case, 
the two overwhelmingly dominant possibilities are young stars and extragalactic 
sources.\footnote{However, we cannot fully rule out the possibility that other
objects might contaminate the sample (a compact planetary nebula, for instance).} We
searched the literature for
previous radio detections, and for counterparts at X-ray, optical, near- and 
mid-infrared wavelengths. The search was done in SIMBAD, and accessed
all the major catalogs (listed explicitly in the footnote of Table\ \ref{tab:counterparts}). 
We considered
a radio source associated with a counterpart at another wavelength if the separation
between the two was below the combined uncertainties of the two datasets. This was
about 1.0 arcsec for the optical and infrared catalogs, but could be significantly larger
for some of the radio catalogs (for instance, the NVSS has a positional uncertainty of
about 5 arcsec).

We found that only 120 of the sources detected here had previously been reported at radio 
wavelengths (column 7 of Table \ref{tab:counterparts}), while the other 491 are new radio
detections. On the other hand, we found a total of 270 counterparts at other wavelengths. 
In the literature, 18 are classified as field stars, 49 as extragalactic, 1 is classified as both
star or extragalactic in different surveys, 49 are classified as YSOs, 11 are classified as
both YSO and extragalactic, the remaining 143 are unclassified.
Note that 56 sources were previously known at radio wavelengths 
and do not have known counterparts at other frequencies. As a consequence, the 
number of sources that were previously known (at any frequency) is 327, while 284
of the sources in our sample are reported here for the first time.

\section{Discussion}

\subsection{Background Sources}

A large number of the sources detected here must correspond to background quasars 
seen in the direction of our targets. We estimate that number as follows. First, we take 
into account the fact that at the edges of the fields, the sensitivity is lower than at their
centers. Following the equation A11 from Anglada et al.~(1998),
who discuss this effect, we obtain that the rough number of
expected background sources with fluxes above $S_{0,\nu}$ in each field and for
each sub-band are:

$$
\langle N\rangle_{4.5\ GHz}=1.21\left(\frac{S_{0,4.5\ GHz}}{{\rm mJy}}\right)^{-0.75}$$

{and}

$$\langle N\rangle_{7.5\ GHz}=0.35\left(\frac{S_{0,7.5\ GHz}}{{\rm mJy}}\right)^{-0.75}.
$$

Using these formulations and $S_{0,\nu}=5\times\sigma_{{\rm noise}\nu}$,
we expect in our 127 fields a total of 731$\pm$27 background
sources at the 4.5 GHz sub-band and 271$\pm$17 at the 7.5 GHz
sub-band. The range of expected values {  was} estimated assuming
a Poisson distribution. Since in our survey we have considerable
overlap of the fields, in particular at 4.5 GHz, the expected
number of sources at 4.5 GHz can be considered as an upper limit.
From our detected sources not related to YSO or stars we counted
530 and 316 in the 4.5 GHz and 7.5 GHz sub-bands, respectively.
This strongly {  suggests} that most of the unidentified sources reported 
here are background sources. 

To analyze this possibility further, we note that
extragalactic radio sources are expected to show little variability 
(typically $\sim$10\% on a timescale of one year) and to have 
negative spectral indices. We have used these characteristics to try 
and separate background sources from YSO candidates in previous 
studies (e.g., Dzib et al. 2013a; Kounkel et al. 2014 and 
Ortiz-Le\'on et al. 2014 {  submitted}). However, there are well known exceptions of 
extragalactic radio sources that do not follow the typical behavior 
(e.g., Heeschen {  1984} and O'Dea 1998) and that we now consider in more
{  detail}. 

Lovell et al.~(2008) found that around half of the flat spectrum
extragalactic sources at centimeter wavelengths {  show} significant
variability on the scale of days to hours. These variations are
attributed mainly to interstellar scintillation by material in our
own Galaxy (e.g., Heeschen \& Rickett 1987; Lovell et al. 2008 and
Koay et al. 2011). We looked through the data of Lovell et al.~(2008)
and found that less than 0.6\% of their sources exhibit high variability, 
using our variability criteria, between two of their four epochs
separated from one another by four months. Assuming that our
sources follow the same trend would indicate that only three of
our background sources should be highly variable. Thus, high 
variability is a good indicator to separate the YSO population from
the background sources.

On the other hand, Gigahertz Peaked-Spectrum (GPS) and High 
Frequency Peakers (HFP) are extragalactic sources that have convex 
radio spectra peaking at frequencies up to 5 GHz for GPS sources 
(see O'Dea 1998
for a review) and above 22 GHz for HFP (Dallacasa et al. 2000).
Thus, the spectral index for these sources in the observed 
frequencies can be flat or positive. About 10\% of the entire
population of radio-bright background sources are GPS, while
20\% of these have {  their} peak above 4 GHz (see O'Dea 1998).
Finally, the analysis by Dallacasa et al. (2000) shows that 
roughly 3\% of their initial sources are HFP. Using these results
we expect that around 21 of the background sources will have
flat or positive spectral indices. Thus, spectral index is
also a useful quantity to separate YSOs from background sources,
but it is somewhat less reliable than variability.

In our sample of 530 possible background sources,  50 are either 
highly variable ($3\times\sigma_{\rm Var.}$ level) or have a flat or positive
spectral index value ({  whitin the range given by 
$\alpha\pm\sigma_{\alpha}$ }). Four of these were previously 
known to be extragalactic sources. The remaining 46 sources are listed 
in Table \ref{tab:candidates}. From the previous arguments, we expect 
about 24 extragalactic sources with these characteristics. Thus, about
half of the sources in Table \ref{tab:counterparts} are most certainly
extragalactic, while the nature of the other half is unclear. They might
be previously unidentified YSOs, but (given the inherent uncertainties 
on the radio properties of different classes of AGNs) they could {  also
 be} extragalactic background sources. 

\subsection{Sources Identified as both YSOs and Extragalactic}

We mentioned earlier that 60 of the detected radio sources are
related to objects previously classified as YSOs. Eleven of these were 
also identified as extragalactic in the Sloan Digital Sky Survey (SDSS, 
Adelman-McCarthy et al.~2011), they are listed in
Table \ref{tab:ysoe}. The radio properties of nine of these
are consistent with being YSO (variability $> 50$\% or
$\alpha$ flat or positive). 

One way to distinguish Galactic from extragalactic sources is by
measuring their proper motions. As extragalactic objects are
far away we do not expect to measure proper motions in them.
On the other hand, members of the Taurus-Auriga region exhibit
proper motions of $\sim20$ mas yr$^{-1}$ (see e.g., Torres et al.~2009).
The proper motions of two of the sources in Table \ref{tab:ysoe}, 
have been measured at radio wavelengths, DG Tau (Rodr{\'{\i}}guez et 
al.~2012a) and DG Tau B (Rodr{\'{\i}}guez et al.~2012b), and were 
found to be consistent with those measured to other members in
Taurus-Auriga.

By combining the position for Haro~6--10~S reported by {  Reipurth et
al.~(2004)}, with those obtained by us, we can roughly estimate 
the proper motions of this source. We obtained $\mu_\alpha \cos{\delta}=16.1\pm7.9$
mas~yr$^{-1}$ and $\mu_\delta=-33.8\pm9.5$ mas~yr$^{-1}$, which
is consistent with it being a member of Taurus-Auriga. 

On the other hand, two of the sources in Table \ref{tab:ysoe} (GBS-VLA J043229.46+181400.2
and GBS-VLA~J041833.38+283732.2) have radio
properties (low variability with negative spectral index) more typical of 
extragalactic objects than of YSOs.

GBS-VLA J043229.46+181400.2  has a radio flux of order $\sim$60 mJy 
which is large compared with most of the other YSOs in Taurus-Auriga. It 
was classified as a Class II YSO
by Gutermuth et al.~(2009) based on its infrared properties. 
We looked in the literature for a good determination of its position
at radio frequencies, but we did not find any. Thus, we looked for a
high resolution, good sensitivity observation in the VLA
archive. We found one observation obtained in 1995 September as part of
the project BB0051. After a standard calibration of these data we
measured the position for this source to be 
RA=04$^{\rm h}$32$^{\rm m}$29\rlap{$^{\rm s}$}.460 and 
Dec=18$^{\circ}$14$'$00\rlap{$''$}.\,19, with an error in position 
of $\sim$0\rlap{$''$}.\,01. Combining this with our own observations, 
we obtain a rough estimate of the proper motions for this source: $\mu_\alpha \cos{\delta}=4.6\pm3.3$ mas~yr$^{-1}$ and 
$\mu_\delta=5.2\pm3.9$ mas~yr$^{-1}$. These results are consistent with
no motion at all at the 1.4$\times\sigma_{\mu}$ level. This strongly {  suggests}
that GBS-VLA J043229.46+181400.3 is a background source rather than a
YSO, and we will not consider it as a YSO in the rest of the paper.

The remaining source, GBS-VLA~J041833.38\\+283732.2, was only observed in the
4.5 GHz sub-band and does not show significant levels of variability.
At infrared wavelengths it was classified as a Class~I YSO, and low
variability is expected for most of the stars in this evolutionary
status (e.g., Dzib et al.~2013a). Also, we note that all the YSOs
in the evolutionary Class~I that are detected by the SDSS, are
classified as extragalactic by them, suggesting a bias in the SDSS 
identification criteria. Thus, we will consider it as a YSO. In summary,
we consider that 10 of the 11 sources in Table \ref{tab:ysoe} are 
counterparts of YSOs, and that 59 of the 60 sources previously 
classified as YSOs are in fact YSOs.

\subsection{General Radio properties of the YSO population}

Sources identified as YSOs are listed in Table \ref{tab:yso}.
From their radio properties, 33 objects show at least one clear indication
that their emission is non-thermal in nature (high variability, 
circular polarized emission or negative spectral index). Another 22 objects
are consistent with {  their} emission being thermal free-free emission
(positive spectral index and low variability). From the
remaining sources three are {  non-variable}, do not show signs of circular 
polarization, but they show a negative spectral index, that given the involved 
errors, the nature of their emissions {  remains} uncertain.
Finally, we do not have enough information to favor a thermal or non-thermal 
emission mechanism for GBS-VLA J041833.38+283732.2. {  Our results show that 
 a significant population of the YSOs in Taurus-Auriga (56\%)  have radio emission
that is non-thermal in nature}. These
stars can be used for VLBI astrometric observations (e.g., Dzib et al.~2010; 
2011 and Torres et al.~2012).

The evolutionary status is known for 54 of the 59 YSOs detected in this 
radio survey. In addition, 53 have a known spectral type. Thus, we can 
compare their radio properties to their evolutionary status and spectral 
type. 

\begin{figure*}[t!]
\begin{center}
\includegraphics[width=0.90\textwidth,angle=0]{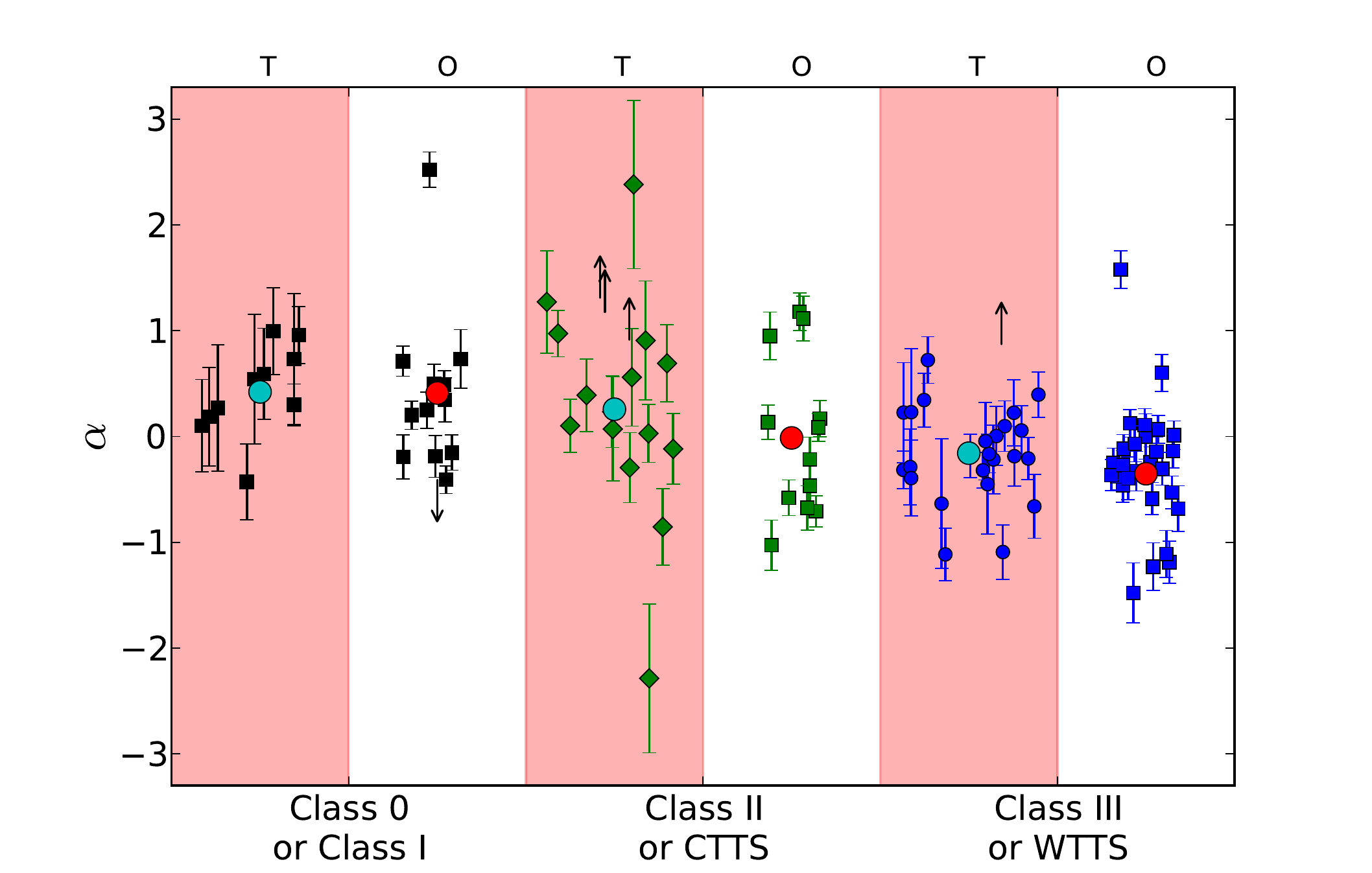}
\end{center}
\caption{Spectral index as a function of the YSO evolutionary stages for YSOs 
in Taurus-Auriga (labeled in the top with a T) and Ophiuchus (labeled in the top with an O).
The individual sources
are shown with their error bars, and the turquoise and red circles indicate the mean spectral
index for each category in Taurus-Auriga and Ophiuchus, respectively. Ophiuchus data from Dzib
 et al.~(2013a).}
\label{fig:ACl}
\end{figure*}

\begin{figure*}[t!]
\begin{center}
\includegraphics[width=01.0\textwidth,angle=0]{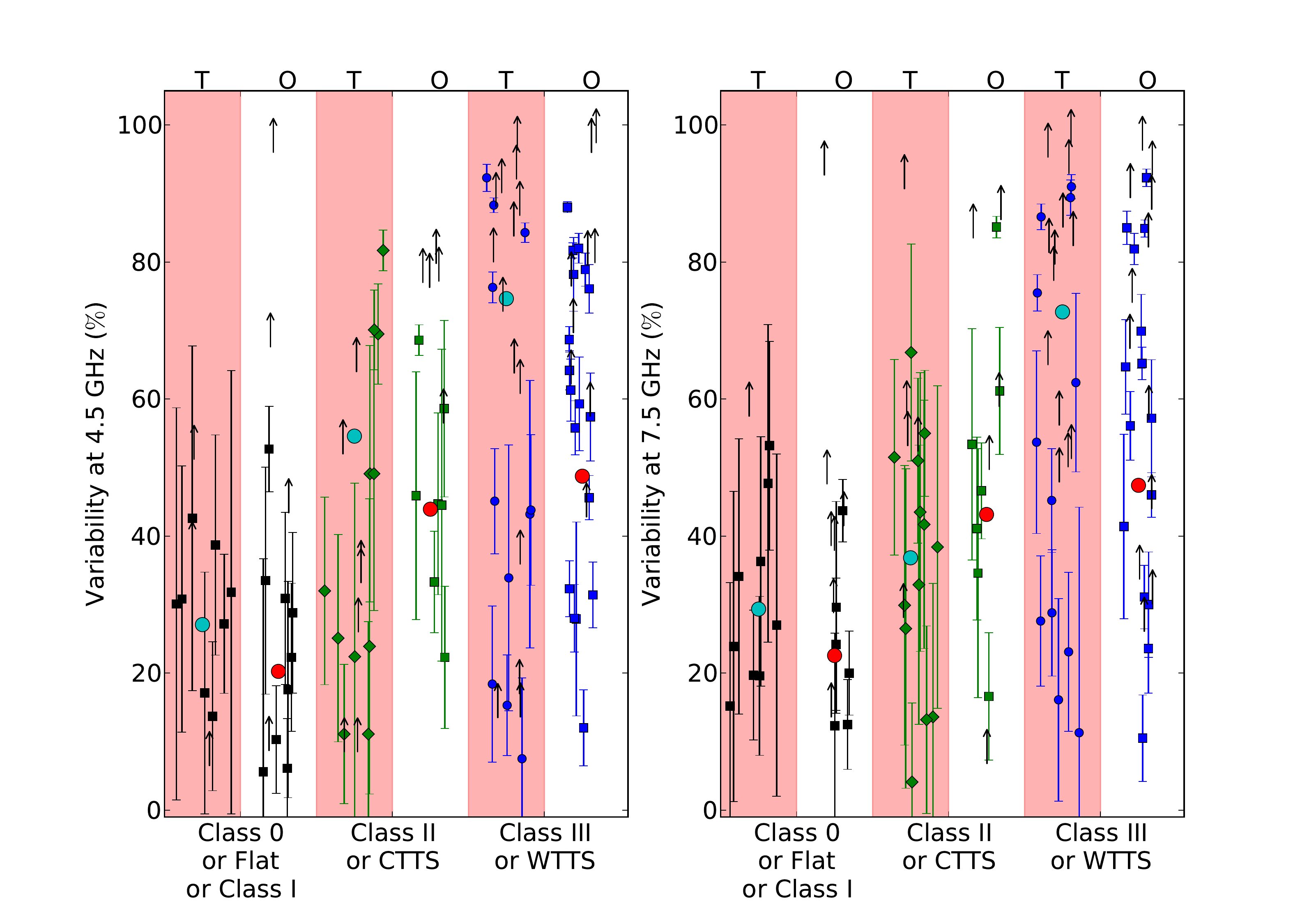}
\end{center}
\caption{Variability at 4.5 GHz (left) and 7.5 GHz (right) as a function of YSO 
evolutionary status. We have labeled in the top with a T and an O if the sources 
are in Taurus or Ophiuchus, respectively.  The individual sources are shown with their error bars, 
and the turquoise and red circles indicate  the mean variability for each category
in Taurus-Auriga and Ophiuchus, respectively. Ophiuchus data from Dzib
 et al.~(2013a).}
\label{fig:VCl}
\end{figure*}

\begin{figure*}[t!]
\begin{center}
\includegraphics[width=01.0\textwidth,angle=0]{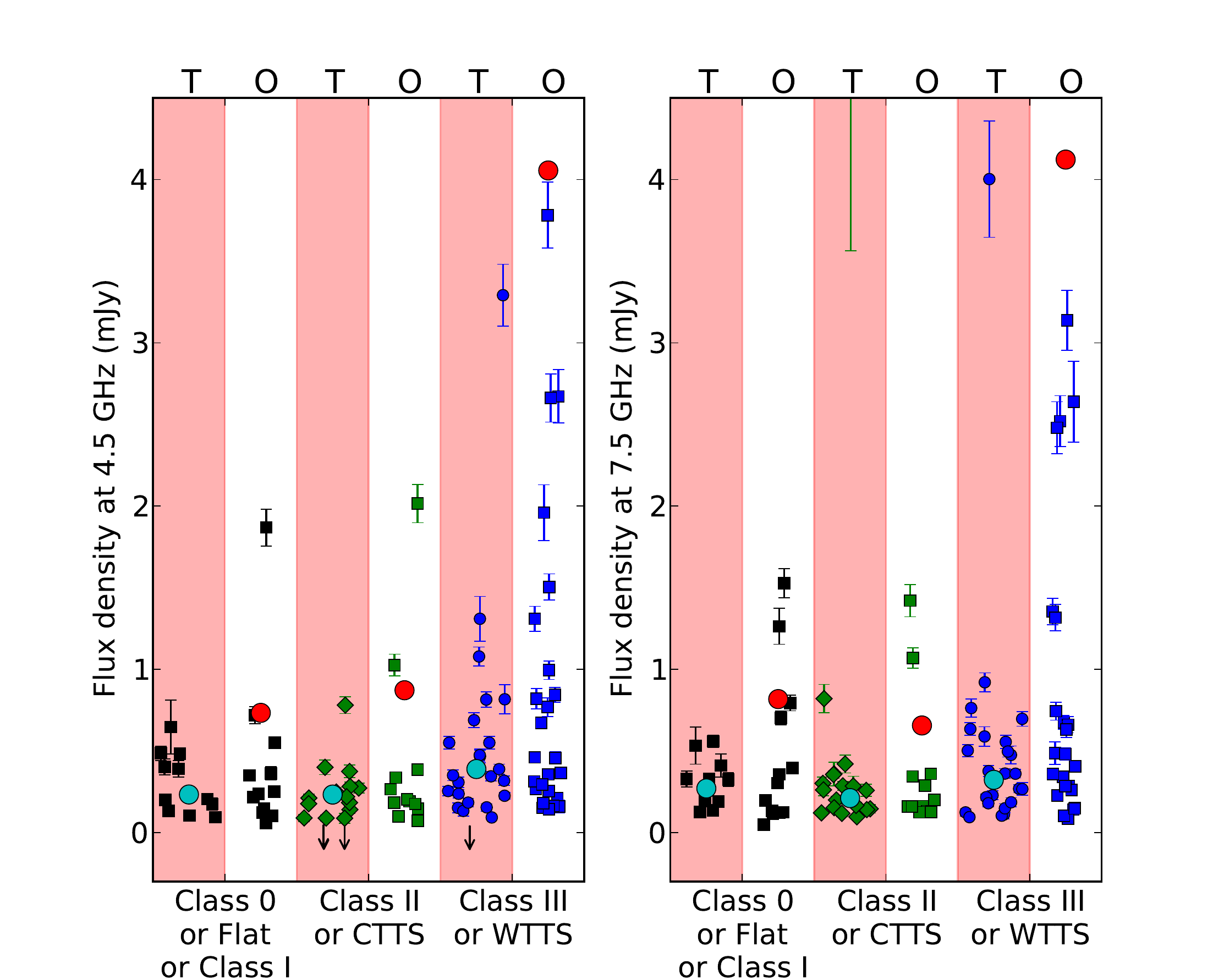}
\end{center}
\caption{Radio flux at 4.5 GHz (left) and 7.5 GHz (right) as a function of YSO 
evolutionary status. We have labeled in the top with a T and an O if the sources 
are in Taurus or Ophiuchus, respectively. As in the previous figures, the individual sources are shown 
with their error bars, and the turquoise and red circles indicate the mean 
flux for each category in Taurus-Auriga and Ophiuchus, respectively.
Ophiuchus data from Dzib et al.~(2013a). We limit the figure to a flux of 4.5 mJy to see the slope 
between the mean fluxes in Taurus.}
\label{fig:FCl}
\end{figure*}

We have found previously that, on average, more evolved stars show
radio properties that resemble a non-thermal origin (i.e., more
variable and with more negative spectral indices, Dzib et al. 2013a;
Ortiz-Le\'on et al. 2014 submitted; and Kounkel et al. 2014). This is
in agreement with the idea that more evolved sources have {  already} shed away most of
their surrounding material and that we are detecting the non-thermal
emission from their coronae instead of the thermal emission from the
surrounding material (see Dzib et al. 2013a for a discussion).
From Figures \ref{fig:ACl} and \ref{fig:VCl} (where we have also plotted our
previous result in Ophiuchus) we see that the YSOs in
Taurus-Auriga follow the same tendency. Note, however, that {  at each
of the evolutionary stages we found} both thermal and non-thermal radio
emitters (see also Table \ref{tab:yso}) as also noted in Ophiuchus (Dzib
et al. 2013a), Serpens and W40 
(Ortiz-Le\'on et al. submitted) and Orion (Kounkel et al. 2014).
The weighted average flux density seems also to be slightly higher
for the more evolved YSOs (see Figure \ref{fig:FCl}), but it is not as
dramatic as in the Ophiuchus case (Dzib et al. 2013a; Figure \ref{fig:FCl}). 

\begin{figure*}[t!]
\begin{center}
\includegraphics[width=01.0\textwidth,angle=0]{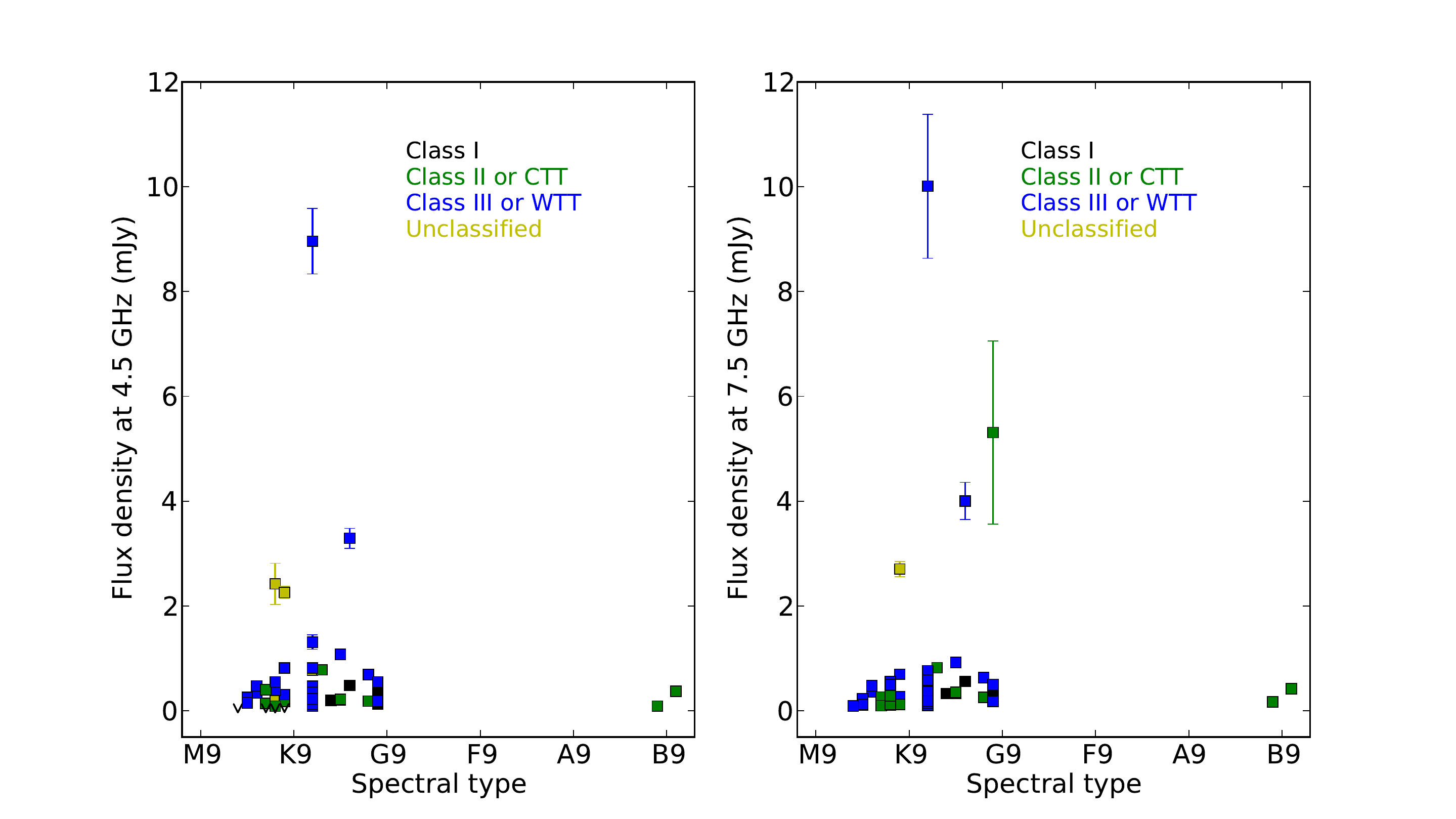}
\end{center}
\caption{Radio flux at 7.5 GHz (left) and 4.5 GHz (right) as a function of YSO 
spectral type. Colors indicate the evolutionary class of the 
object as listed at the top-right of the diagrams.}
\label{fig:FST}
\end{figure*}

Finally, Figure \ref{fig:FST} shows the flux density of the YSOs as a
function of their spectral type. As in the Ophiuchus case, most
of the YSOs sources detected at radio frequencies have a spectral type 
M and K, there are no detected sources of types G and F, and only one 
A-type and one B-type detection. In part, this presumably reflects the fact 
that the Taurus-Auriga star forming-region {  contains mainly low mass stars}. 
Contrary to the results in Ophiuchus, in Taurus-Auriga the averaged radio 
flux density for the low-mass members seems to be larger than the 
intermediate-mass members. We discuss the peculiarities of these stars in 
the next section.

\subsection{The radio -- X-ray relation}

An interesting result that we have been noticing in our VLA
surveys is the relation between the radio emission of YSOs
compared to their previously reported X-ray emission. The
recent results indicate that YSOs follow the so-called
G\"udel-Benz relation (Guedel \& Benz~1993 and Benz \& 
Guedel~1994) for magnetically active stars. The relation
has the general form (Benz \& Guedel~1994):

\begin{displaymath}
 \frac{{\rm L}_X}{L_{R,\nu}}=\kappa\cdot10^{15.5\pm0.5}\quad{\rm [Hz]}.
\end{displaymath}

The value of $\kappa$ varies for different kind of stars and fulfills
$\kappa\leq1$. Particularly, we proposed a $\kappa=0.03$ for YSOs, based
just in the positions on the L$_X$--L$_R$ diagram of YSOs in the Ophiuchus
star forming regions (Dzib et al. 2013a), and also following the previous
results obtained by Gagn\'e et al. (2004) and Forbrich et al. (2010).

\begin{figure*}[t!]
\begin{center}
\includegraphics[width=0.60\textwidth,angle=0]{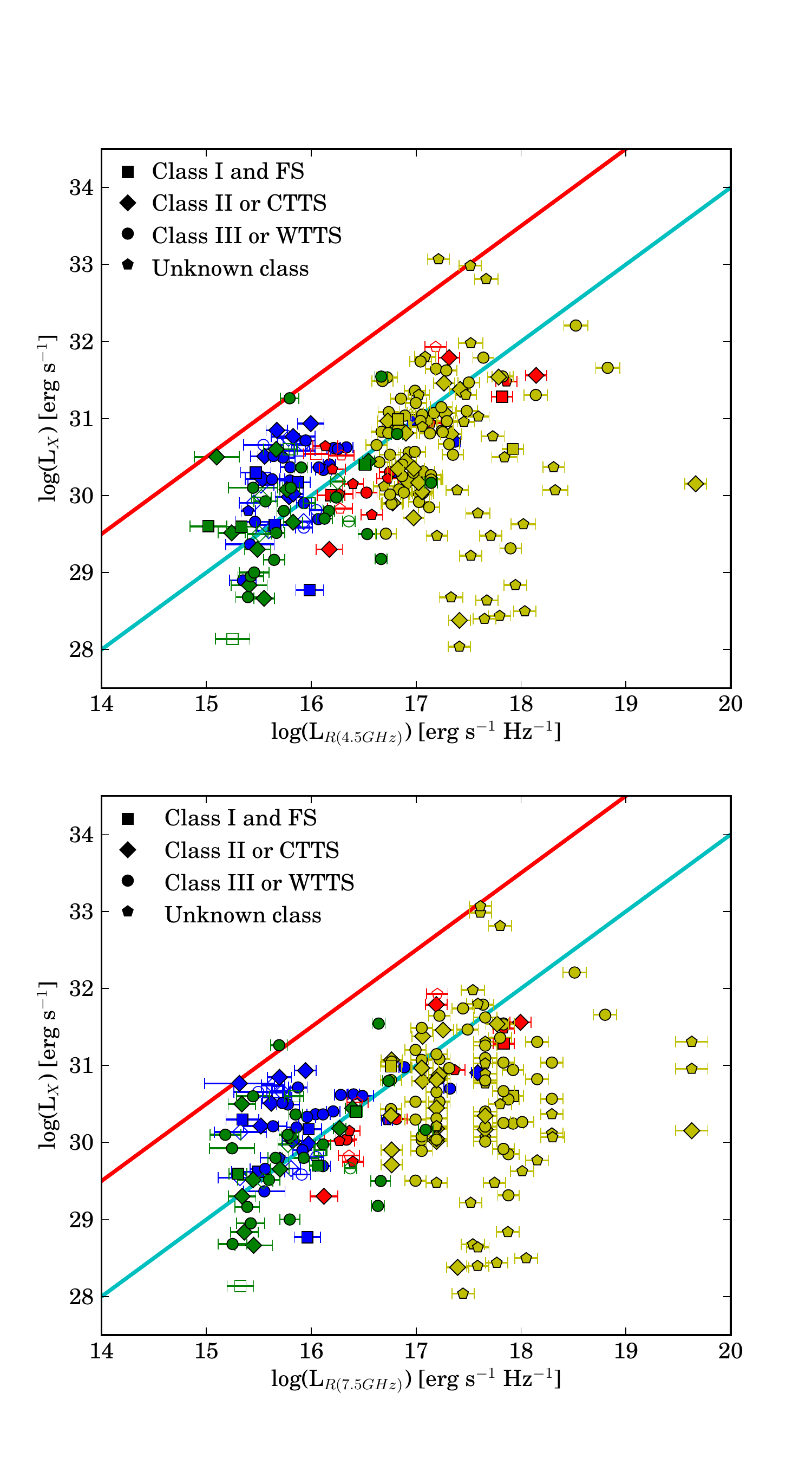}
\end{center}
\caption{\footnotesize{X-ray luminosity as a function of radio luminosity. The red line corresponds to the 
G\"udel-Benz relation with $\kappa=1$. The blue line correspond to
the G\"udel-Benz relation but with $\kappa=0.03$. Symbols indicate the evolutionary status 
of the object as explained at the top-left of the diagram. Colors indicates YSOs in different
star forming regions: Taurus (blue, this work), Ophiuchus (green, Dzib et al.~2013a),
Serpens-W40 (red, Ortiz-Le\'on et al.~2014, submitted) and Orion (yellow, Kounkel et al.~2014). Open symbols indicate
sources whose radio emission is thermal and solid symbols indicates non-thermal radio sources.}
}
\label{fig:GB}
\end{figure*}

We searched for X-ray counterparts of the YSOs detected in
Taurus-Auriga (see also Table \ref{tab:counterparts}) and plot
them against the radio luminosities in Figure \ref{fig:GB} with
blue symbols. In Figure \ref{fig:GB} we also plotted the
results obtained in Ophiuchus (green symbols, from Dzib et
al.~2013a), Serpens-W40 (red symbols; from Ortiz-Le\'on et
al.~submitted), and Orion (yellow symbols; from Kounkel et al.
2014). Most of the YSOs seem to follow, with $\sim$1~dex of
dispersion, the G\"udel-Benz relation with $\kappa=0.03$.
There are, however, some clear exceptions in the
Orion region, where some sources are apparently X-ray 
under-luminous by more than two orders of magnitude.
We noticed that this trend does not
depend on the evolutionary classification of the YSOs 
(squares for Class~I/FS, rhombus for Class~II/CTTS, circles
for Class~III/WTTS and pentagons for unclassified sources
in Figure \ref{fig:GB}) or if the radio emission is most likely to be thermal
(open symbols) or non--thermal (filled symbols). Dzib et al.~(2013b)
has discussed that even when the gyrosynchrotron radio emission
is hidden by free-free radio emission in upper layers of circumstellar
material, the X-ray emission produced in the plasma of the corona
could {  still} be observed. Even more interesting is that, as noticed by Dzib
et al.~(2013b), independent of the dominant radio emission mechanism the YSOs
seem to fit the G\"udel-Benz relation, meaning that both radio emission
mechanisms {  have} similar levels of flux densities.

\section{Comments on some individual sources}

\subsection{Non-thermal Class~I Stars}

{  Newly-born} low-mass stars (between 0.08 M$_\odot$ to 3.0 M$_\odot$) are
magnetically active and could produce significant amount of gyrosynchrotron
radio emission from their {  coronae}. In the very young stars (i.e., Class I) this radio
emission is expected to be hidden by the optically thick free-free
radio emission from the surrounding material. There are a few cases,
however, where non-thermal radio emission has been detected from Class I 
stars (e.g., Feigelson et al. 1998; Forbrich et al. 2007; Dzib et al. 
2010, 2013a and Deller et al. 2013). Geometric effects or abnormally early
removal of the surrounding material (e.g.\ in a tight binary system) have been 
invoked to explain these detections (e.g., Forbrich et al. 2007; Dzib et al. 2010).
Thus, it is important to document more cases  in order to identify
which is the dominant mechanism that allows the detection of coronal emission
in very young YSOs. Three Class~I stars in our sample have properties {  that indicate 
their radio} emission could be of a non-thermal nature:
{\it GBS-VLA J041354.72+281132.6}, {\it GBS-VLA J043232.07+225726.3}
and {\it GBS-VLA J0439\\35.20+254144.3} are  Class~I stars that are highly
variable, which is an indication that their emission has a 
non-thermal origin. In the case of {\it GBS-VLA J043232.07+225726.3}
this is also supported by the negative spectral index.
The non-thermal origin of the emission from these very young stars must be confirmed using
other techniques, e.g, Very Long Baseline Interferometry (VLBI).

\subsection{Low Variability Non-thermal Radio Emission}

Non-thermal radio emission from YSOs often exhibits high levels
of variability ($>$50\%) asociated with magnetic activity 
of the star (Feigelson \& Montmerle 1999). There are, however,
some exceptions to this. The 6 M$_\odot$ star S1 in the Ophiuchus
core shows only modest levels of variability during
a series of VLBI observations made by Loinard et al.~(2008).
This suggests a different origin with more organized magnetic fields.
For a better understanding of this behavior, further studies must
be carried with more objects. In the present report we detected 
two other sources which are likely non-thermal because they have
detectable circular polarization, but which are only moderately variable.

GBS-VLA J041831.12+282715.9 = V~410~Tau, is a K4 star that shows variabilities
below 30\% during the three observed epochs. Circular polarized emission
was found at a level of 8.7\% (R) in the 7.5 GHz sub-band. Also, it shows
a negative spectral index ($\alpha=-0.32\pm0.17$) which supports the
non-thermal origin. Carroll et al.~(2012) found that the magnetic
topology in the visible pole of V~410 is dominated by a bipolar structure.
These authors also argue that the absence of a radiative core and non-detection
of differential rotation support the idea that a classical dynamo is
not operating in V~410~Tau.

GBS-VLA J042203.15+282538.8 is an M3  star with measured variability
below 16\%. Its spectral index between the observed sub-bands is flat 
($\alpha=0.00\pm0.28$). Circular polarization is detected in both sub-bands
at levels of 21.3\% (L) and 30.8\% (L) for 4.5 GHz and 7.5 GHz, respectively.
These levels of circular polarization strongly suggest a non-thermal origin.

\subsection{Sources with large positive spectral indices}

Only two sources have spectral indices that are consistent with large
positive values ($\sim$2). These sources are GBS--VLA J043540.95+241108.6 
(with spectral index 2.38$\pm$0.79) and GBS--VLA
J0440\\22.18+260515.2 (with spectral index $>$1.70$\pm$0.14).
The first source is associated with CoKu Tau 3, a well known Class II
YSO and we attribute the spectral index to the presence of optically-thick
gyrosynchrotron emission. The second source, GBS--VLA J044022.18+260515.2,
has been tentatively classified by us as a YSO candidate on the basis of
its positive spectral index. An alternative interpretation is that it is
a High Frequency Peaker (HFP).
These are compact extragalactic radio sources with well-defined peaks in their
radio spectra above 5 GHz and positive spectral indices at frequencies below their
peak. Most of them are believed to be high redshift quasars (Orienti \& Dallacasa 
2014). The case of a possible HFP observed in projection in the star forming 
region M17 is discussed by Rodr\'{\i}guez et al. (2014a).

\subsection{The very low-mass star GBS-VLA J043158.46+254329.8}

Dzib et al. (2013a) show that the detection of YSOs near and beyond
 the brown dwarf boundary at distances farther than 100 pc is
possible. {  They reported the radio emission of four very
low-mass stars in the Ophiuchus core.} Three of them show indications that the
origin of their radio emission is non-thermal.

GBS-VLA J043158.46+254329.8 is a Class~III YSO (Kenyon \& Hartmann 1995), 
with a mass of 0.19 M$_\odot$ and an M5.5 spectral type (Brice\~{n}o et al.~2002).
It was not detected in the 4.5 GHz sub-band on individual epochs at levels
of 100~$\mu$Jy nor in the image of the three concatenated epochs
at a level of 63~$\mu$Jy. On the other hand, in the 7.5 GHz sub-band it was not
detected in the first two epochs at levels of 80~$\mu$Jy, but in the third epoch
it was detected with a total flux density of 165$\pm28~\mu$Jy. This variation 
($>$45\%) is just slightly below our adopted limit of high variability
and suggests that GBS-VLA J043158.46+254329.8 could also be a very low-mass
star with non-thermal radio emission.

\subsection{Non-Thermal Radio Emission from the Close Binary Elias~1=GBS-VLA J041840.62+281915.3}

Elias~1 is a well known Herbig AeBe (HAeBe)\footnote{HAeBe stars
are pre-main sequence intermediate mass stellar objects, i.e.,\ the intermediate
mass analogs of T Tauri stars.}  star in the Taurus-Auriga cloud (Elias 1978).
It has two companions, a T Tauri star at 4$''$ to its north-east (Leinert et
al.1997) and a close (50 milli-arcsec) low-mass companion (Smith et al.~2005).
It is variable at X-rays frequencies, and an X-ray flare was reported by
Giardino et al.~(2004). This is an indication that a corona is acting in at least 
one of the two components of the close binary. At radio frequencies, we detected
a source related with the close binary (GBS J041840.62+281915.3). It is a
highly variable star in both of the observed sub-bands (69.5$\%\pm16.7$\% and
 $51.5\%\pm15.2\%$ at 4.5 GHz and 7.5 GHz, respectively). This is another indication
that a corona is involved. 

The standard theory of intermediate-mass stellar evolution predicts that during
the pre-main sequence (and, indeed, the main sequence), these stars 
of relative large masses are fully
radiative. As a consequence, they are not expected to drive a dynamo, and should
not have strong surface magnetic fields. Thus, they should not maintain a corona. 
X-ray observations have shown that some HAeBe stars have indeed signs that they
can host {  coronae} that resemble those of T Tauri stars (Damiani et al.\ 1994; 
Zinnecker \& Preibisch 1994; Stelzer et al.\ 2005; Stelzer et al.\ 2006 and Stelzer
et al.\ 2009). On account of that similarity, a popular interpretation was that 
the X-ray emission was in fact due to the presence of a young low-mass stellar
companion (rather than to the HAeBe star itself). {  The hot X-ray component
of the single HAeBe star HD 163296, on the other hand, suggested a coronal origin
and is the most promising case of an HAeBe hosting a corona (G{\"u}nther \& 
Schmitt 2009).} To our knowledge, non-thermal
radio emission from young intermediate mass stars has only been detected directly,
using the VLBI technique, from the B4V S1 star in the Ophiuchus cloud (Andre et 
al.~1991; Loinard et al.~2008) and from the proto-HAeBe binary star EC~95 
in the Serpens cloud (Dzib et al.~2010).
The high variability of EC~95 at radio frequencies strongly suggests that it has a
coronal origin (see Dzib et al.~2010 for a discussion), while the low variability 
in S1 suggests a different origin. However, a corona has never been
directly observed from a well established HAeBe star. Future VLBI observations of GBS-VLA
J041840.62+281915.3 will establish if the non-thermal radio emission observed with the VLA
comes from the HAeBe star itself or from its low mass companion. Even the detection
of the companion with the VLBI technique is of  great interest because we will be able 
to estimate the dynamical mass of the HAeBe star.

\subsection{GBS-VLA J045545.85+303304.0 = AB Aur}

The lack of variability and spectral index (1.27$\pm$0.48) of this Herbig Ae star 
suggests that the emission is produced by a thermal jet. This suggestion is
confirmed by the study of Rodr\'\i guez et al. (2014b)
that combines additional VLA data at other wavelengths  with our results.



\section{Conclusions and perspectives}

We presented a multi-epoch VLA survey at two frequencies of one
of the closest and best studied star forming region, the
Taurus-Auriga cloud. It is more sensitive, covers a larger region, 
and has higher angular resolution than all previous surveys of this
region. The multi-epoch, two frequency strategy {  has enabled} 
us to determine the radio properties of the
detected sources, and {  has provided clues about the nature of their radio
emission and the nature of the objects}. We detected a
total of 610 radio sources, 59 of them related to YSOs and 18 to field stars.
We argue that most of the remaining objects are extragalactic
sources, but we provide a list of sources whose radio characteristics would
be consistent with their being YSOs. Up to half of them {  may} be 
previously unidentified YSOs. 

The radio emission for most of the YSOs (56\%), is consistent with
a non-thermal origin (gyrosynchrotron); these sources could be used for future
VLBI observations. In line with our studies of other star-forming 
regions like Ophiuchus, Serpens and Orion (Dzib et al. 2013a; 
Ortiz-Le\'on et al. 2014 {  submitted}, and Kounkel et al. 2014), we find in Taurus
that the radio emission tends to be of a more non-thermal nature
for {  the} more evolved YSOs. Finally, by comparing our results
with previous X-ray observations we obtained that the sources in
Taurus-Auriga follow the so-called G\"udel-Benz relation with 
$\kappa=0.03$, however with a somewhat large dispersion, also 
consistent with the results in other star forming regions.

\acknowledgments
L.L., L.F.R., G.N.O., G.P., and J.L.R. acknowledge the financial 
support of DGAPA, UNAM, and CONACyT, M\'exico. 
NJE was supported in part by
NSF Grant AST-1109116 to the University of Texas at Austin.
The National Radio Astronomy Observatory 
is operated by Associated Universities Inc. under cooperative agreement with the National 
Science Foundation. CASA is developed by an international consortium of 
scientists based at the National Radio Astronomical Observatory (NRAO), the European 
Southern Observatory (ESO), the National Astronomical Observatory of 
Japan (NAOJ), the CSIRO Australia Telescope National Facility (CSIRO/ATNF), 
and the Netherlands Institute for Radio Astronomy (ASTRON) under the 
guidance of NRAO. This research has made use of the SIMBAD database,
operated at CDS, Strasbourg, France

\newpage
\begin{deluxetable}{ccrcrc}
\tabletypesize{\scriptsize}
\tablewidth{0pt}
\tablecolumns{6}
\tablecaption{Radio Sources Detected in Taurus-Auriga. \label{tab:sources}}
\tablehead{           & \multicolumn{4}{c}{Flux Properties}	&Spectral\\
\colhead{GBS-VLA Name} & \colhead{$f_{4.5}$(mJy)\tablenotemark{a}} &\colhead{Var.$_{4.5}$\,(\%)} & \colhead{$f_{7.5}$(mJy)\tablenotemark{a}} & \colhead{Var.$_{7.5}$\,(\%)}&\colhead{Index}\\
}\startdata
J040331.11+260911.4&(1.45$\pm$0.16$\pm$0.07)$\times10^{+0}$&19.8$\pm$15.0&\nodata&\nodata&\nodata\\
J040345.47+261612.6&(5.54$\pm$1.17$\pm$0.28)$\times10^{-1}$&64.1$\pm$13.6&\nodata&\nodata&\nodata\\
J040349.35+261051.8&(3.45$\pm$0.23$\pm$0.17)$\times10^{-1}$&33.9$\pm$19.4&(3.62$\pm$0.24$\pm$0.18)$\times10^{-1}$&23.1$\pm$11.6&0.10$\pm$0.24\\
J040407.97+261402.1&(7.72$\pm$1.03$\pm$0.39)$\times10^{+0}$&\nodata&\nodata&\nodata&Extended\\
J040416.39+261544.5&(2.02$\pm$0.33$\pm$0.10)$\times10^{+0}$&36.2$\pm$17.6&\nodata&\nodata&\nodata\\
J040434.84+261810.7&(2.36$\pm$0.34$\pm$0.12)$\times10^{-1}$&35.7$\pm$21.8&(1.62$\pm$0.46$\pm$0.08)$\times10^{-1}$&$>$29.4$\pm$27.5&-0.76$\pm$0.66\\
J040443.07+261856.3&(2.06$\pm$0.20$\pm$0.10)$\times10^{-1}$&17.1$\pm$17.6&(3.31$\pm$0.19$\pm$0.17)$\times10^{-1}$&19.6$\pm$11.6&0.96$\pm$0.27\\
J040505.66+262311.5&(1.80$\pm$0.35$\pm$0.09)$\times10^{+0}$&44.2$\pm$16.2&\nodata&\nodata&\nodata\\
J041305.34+282255.1&(2.77$\pm$0.45$\pm$0.14)$\times10^{-1}$&35.9$\pm$19.0&(3.26$\pm$1.77$\pm$0.16)$\times10^{-1}$&\nodata\tablenotemark{b}&0.33$\pm$1.15\\
J041305.40+281413.9&(2.28$\pm$0.47$\pm$0.11)$\times10^{-1}$&22.8$\pm$28.9&\nodata&\nodata&\nodata\\
\enddata
\tablecomments{Table \ref{tab:sources} is published in its entirety in the electronic edition of the {\it Astrophysical Journal}. A portion is shown here for guidance regarding its form and content.}
\tablenotetext{a}{ The two given errors correspond to (i) statistical noise in the images and (ii) systematic uncertainty from possible errors in flux calibration.}
\tablenotetext{b}{Source not detected at three times the noise level on individual epochs, but detected on the image of the concatenated epochs.}
\end{deluxetable}

\begin{deluxetable}{cccc}
\tabletypesize{\scriptsize}
\tablewidth{0pt}
\tablecolumns{6}
\tablecaption{Sources detected in circular polarization}
\tablehead{{GBS-VLA Name} & Source type & 4.5 GHz polz (\%) & 7.5 GHz polz (\%)\\}
\startdata
J041327.23+281624.4 & YSO/Class~II      & 36.3 (L)      &   $<$30.5 \\
J041628.11+280735.4 & YSO/Unknown       & 9.6 (L)      &   18.8 (L) \\
J041831.12+282715.9 & YSO/Class~III     & $<$6.5       &   8.7 (R) \\
J041840.62+281915.3 & YSO/Class~II      & 32.0 (L)     &   $<$29.9 \\
J041847.04+282007.2 & YSO/Class~III     & 3.1 (L)      &   5.6 (L) \\
J041941.28+274947.9 & YSO/Unknown       & 5.1 (L)     &   $<$2.6 \\
J042159.43+193205.7 & YSO/Class~II      & 69.7 (L)    &   \nodata \\
J042203.15+282538.8 & YSO/Class~III     & 21.3 (L)    &   30.8 (L) \\
J043140.09+181356.7 & YSO/Class~III     & 35.8 (L)    &   31.2 (L) \\
J043214.58+182014.6 & YSO/Class~III     & 17.8 (R)    &    $<$6.6\\
J043542.05+225222.4 & Star              & 28.8 (R)    &    \nodata\\
\enddata
\label{tab:polz}
\end{deluxetable}

\begin{deluxetable}{lccccccc}
\tabletypesize{\scriptsize}
\tablewidth{0pt}
\tablecolumns{8}
\tablecaption{Radio Sources with known counterparts. \label{tab:counterparts}}
\tablehead{           & Other         &         & \multicolumn{3}{c}{Infrared\tablenotemark{b}} &       & Object\\
\colhead{GBS-VLA Name} & \colhead{Names} &\colhead{X-ray\tablenotemark{a}} & \colhead{SST} & \colhead{2M}&\colhead{WISE}&
\colhead{Radio\tablenotemark{c}}&\colhead{type}\\}
\startdata
J040349.35+261051.8&2MASS J04034930+2610520&XEST&Y&Y&Y&--&YSO\\
J040416.39+261544.5&WISE J040416.33+261545.2&--&--&--&Y&--&--\\
J040434.84+261810.7&WISE J040434.83+261810.5&--&--&--&Y&--&--\\
J040443.07+261856.3&IRAS 04016+2610&XEST&--&Y&Y&--&YSO\\
J040505.66+262311.5&WISE J040505.68+262311.2&--&--&--&Y&NVSS&--\\
J041305.34+282255.1&WISE J041305.31+282255.1&--&Y&Y&Y&--&--\\
J041305.40+281413.9&--&--&--&--&--&NVSS&--\\
J041305.45+281414.7&--&--&--&--&--&NVSS&--\\
J041314.16+281910.3&V1095 Tau&XEST&Y&Y&Y&--&YSO\\
J041327.23+281624.4&V1096 Tau&XEST&Y&Y&Y&--&YSO\\
\enddata
\tablecomments{Table \ref{tab:counterparts} is published in its entirety in the electronic edition of the {\it Astrophysical Journal}. A portion is shown here for guidance regarding its form and content.}
\tablenotetext{a}{XEST = G\"udel et al. (2007); 1RXS = Voges et al. (1999); RX = Neuhaeuser et al.\,(1995) and 
BFR2003 = Bally et al.\,(2003).}
\tablenotetext{b}{SST = Padgett et al.\ (2007); 2M = Cutri et al.\ (2003) and WISE = Cutri et al.\ (2013).}
\tablenotetext{c}{
NVSS = Condon et al.\ (1998); BW = Becker \& White (1985); SBS = Skinner et al.\,(1993); OFMM~=~O'Neal~et~al.~(1990);
LRDRG = Loinard et al.\,(2007b); RAR = Rodr\'iguez et al.\,(1995); CB = Cohen \& Bieging (1986); 
ARC92 = Anglada et al.\,(1992); RRAB = Reipurth et al.\,(2004); B96 = Bontemps (1996); GRL2000 = Giovanardi et al.\,(2000)
and RR98 = Rodr\'iguez \& Reipurth\,(1998).}
\end{deluxetable}

\begin{deluxetable}{cccc}
\tabletypesize{\scriptsize}
\tablewidth{0pt}
\tablecolumns{4}
\tablecaption{Young Stellar Object Candidates Based Only on Their Radio Properties \label{tab:candidates}}
\tablehead{\colhead{GBS-VLA Name} &\colhead{Variability$_{4.5 {\rm GHz}}$\,(\%)} & \colhead{Variability$_{7.5 {\rm GHz}}$\,(\%)}&\colhead{Spectral Index}\\
}\startdata
J040345.47+261612.6&64.1$\pm$13.6&\nodata&\nodata\\
J041305.45+281414.7&70.2$\pm$13.8&\nodata&\nodata\\
J041334.03+292255.3&$>$60.7$\pm$16.7&\nodata&\nodata\\
J041412.58+281155.7&38.9$\pm$18.0&40.6$\pm$21.0&0.42$\pm$0.50\\
J041437.95+264604.3&13.2$\pm$12.7&39.3$\pm$14.6&1.12$\pm$0.35\\
J041454.56+264716.2&$>$59.9$\pm$15.2&16.1$\pm$38.1&-0.51$\pm$0.55\\
J041515.93+291244.5&$>$68.2$\pm$13.8&\nodata&\nodata\\
J041821.80+282956.3&$>$44.3$\pm$23.9&$>$15.0$\pm$28.3&1.06$\pm$0.90\\
J041825.42+252156.4&47.0$\pm$14.2&$>$76.8$\pm$17.5&-1.53$\pm$1.08\\
J041840.27+282036.0&$>$61.1$\pm$13.9&10.7$\pm$37.2&-0.63$\pm$0.68\\
J041844.85+282804.4&52.2$\pm$16.8&$>$45.9$\pm$29.0&0.17$\pm$0.86\\
J041935.50+290628.4&34.3$\pm$21.7&\nodata\tablenotemark{a}&$<$0.01$\pm$0.14\\
J042117.65+270151.0&31.9$\pm$9.1&52.2$\pm$16.1&-1.23$\pm$0.38\\
J042124.18+270025.2&33.0$\pm$14.6&37.0$\pm$23.4&0.26$\pm$0.42\\
J042216.27+265458.7&41.7$\pm$11.7&53.5$\pm$11.3&-0.36$\pm$0.27\\
J042242.40+264201.8&17.2$\pm$12.2&$>$72.9$\pm$10.3&-0.82$\pm$0.51\\
J042317.57+264228.7&$>$37.3$\pm$24.4&33.5$\pm$23.2&1.11$\pm$0.62\\
J042426.28+261050.1&55.2$\pm$13.9&\nodata\tablenotemark{a}&$<$-0.36$\pm$0.14\\
J042634.81+260556.1&55.0$\pm$10.6&\nodata&\nodata\\
J042704.95+260429.3&51.9$\pm$14.2&41.3$\pm$23.7&0.01$\pm$0.53\\
J042917.12+263023.3&29.6$\pm$13.5&$>$62.3$\pm$16.0&-0.49$\pm$0.61\\
J042920.74+263353.4&65.8$\pm$9.9&46.8$\pm$8.3&0.65$\pm$0.26\\
J042939.80+263222.9&51.2$\pm$16.6&$>$27.2$\pm$29.1&-1.05$\pm$0.49\\
J043055.31+260451.9&13.0$\pm$26.4&\nodata\tablenotemark{a}&$<$-0.05$\pm$0.14\\
J043109.20+271045.3&57.9$\pm$7.7&23.4$\pm$12.0&1.02$\pm$0.26\\
J043142.44+181101.5&$>$55.1$\pm$16.1&\nodata\tablenotemark{a}&$<$-1.68$\pm$0.14\\
J043237.91+242054.5&32.7$\pm$9.7&65.4$\pm$10.1&-0.41$\pm$0.41\\
J043323.81+260326.8&55.7$\pm$16.5&\nodata&\nodata\\
J043635.42+241324.2&50.1$\pm$14.3&$>$38.5$\pm$28.3&-0.43$\pm$0.80\\
J043657.44+241835.1&63.0$\pm$16.4&\nodata&\nodata\\
J043816.85+252338.4&62.1$\pm$14.0&\nodata&\nodata\\
J043925.79+253825.3&61.0$\pm$15.5&$>$28.6$\pm$43.9&0.85$\pm$1.07\\
J044004.86+260421.2&27.4$\pm$19.9&50.4$\pm$13.6&0.08$\pm$0.45\\
J044022.18+260515.2&\nodata\tablenotemark{a}&6.2$\pm$45.3&$>$1.70$\pm$0.14\\
J044042.98+255131.9&33.2$\pm$14.6&1.3$\pm$26.5&0.59$\pm$0.44\\
J044123.47+245528.1&73.5$\pm$9.2&\nodata&\nodata\\
J044144.92+255815.3&$>$69.0$\pm$12.8&57.1$\pm$22.4&-0.43$\pm$0.65\\
J044210.59+252505.4&58.5$\pm$11.7&58.0$\pm$19.9&-0.68$\pm$0.61\\
J044233.26+251440.0&48.4$\pm$13.3&15.8$\pm$23.8&0.44$\pm$0.41\\
J044239.84+252306.4&60.3$\pm$14.6&\nodata&\nodata\\
J044246.20+251806.2&81.5$\pm$6.5&$>$28.6$\pm$39.8&-0.54$\pm$0.84\\
J044256.30+251804.6&27.7$\pm$23.5&49.7$\pm$25.7&0.61$\pm$0.77\\
J045539.28+301627.2&24.4$\pm$9.3&32.2$\pm$17.2&0.49$\pm$0.39\\
J045544.38+302455.1&62.7$\pm$9.0&48.9$\pm$33.1&-0.83$\pm$0.81\\
J045547.39+303434.8&50.2$\pm$9.5&30.5$\pm$23.8&-0.83$\pm$0.47\\
J045607.27+302728.2&37.9$\pm$15.7&$>$30.2$\pm$23.3&0.73$\pm$0.90\\

\enddata
\tablenotetext{a}{Source not detected at three times the noise level on individual epochs, but detected on the image of the concatenated epochs.}
\end{deluxetable}





\begin{deluxetable}{ccccc}
\tabletypesize{\scriptsize}
\tablewidth{0pt}
\tablecolumns{5}
\tablecaption{Radio sources with counterparts that are clasified as both YSOs and Extragalactic.}
\tablehead{\colhead{GBS-VLA Name} &  \colhead{Other Names} &\colhead{Variability$_{4.5 {\rm GHz}}$\,(\%)} & \colhead{Variability$_{7.5 {\rm GHz}}$\,(\%)}&\colhead{Spectral Index}\\
}\startdata
J041833.38+283732.2&2MASS J04183343+2837321&30.1$\pm$28.6&\nodata&\nodata\\
J041851.49+282026.1&CoKu Tau 1&30.8$\pm$19.4&23.9$\pm$22.6&0.27$\pm$0.60\\
J042200.70+265732.2&2MASS J04220069+2657324&31.8$\pm$32.3&27.0$\pm$25.0&0.54$\pm$0.61\\
J042202.20+265730.3&FS Tau A-B             &$>$51.6$\pm$17.3&$>$52.8$\pm$18.7&-0.29$\pm$0.33\\
J042448.16+264315.9&V1201 Tau              &45.1$\pm$7.7&45.2$\pm$7.6&-0.16$\pm$0.18\\
J042702.56+260530.4&DG Tau B               &25.1$\pm$15.1&13.6$\pm$19.5&0.39$\pm$0.34\\
J042704.69+260615.8&DG Tau                 &11.1$\pm$10.2&13.2$\pm$13.7&0.10$\pm$0.25\\
J042923.74+243259.8&HARO 6-10 S             &13.7$\pm$10.9&19.7$\pm$9.5&0.30$\pm$0.20\\
J043138.42+181357.3&HL Tau                 &13.9$\pm$17.6&1.3$\pm$17.1&0.68$\pm$0.36\\
J043229.46+181400.2&2MASS J04322946+1814002&10.2$\pm$14.3&17.2$\pm$38.9&-0.25$\pm$0.71\\
J043439.24+250100.9&V1110 Tau              &$>$63.4$\pm$12.7&$>$76.9$\pm$8.1&-0.04$\pm$0.37\\

\enddata
\label{tab:ysoe}
\end{deluxetable}

\begin{deluxetable}{lcccccccc}
\tabletypesize{\scriptsize}
\tablewidth{0pt}
\tablecolumns{8}
\tablecaption{Young Stellar Objects detected in the radio observations \label{tab:yso}}
\tablehead{           & Spectral       &  & SED &                   &  &  & &  \\
\colhead{GBS-VLA Name} & \colhead{type}&\colhead{Ref.}& \colhead{clasification}&\colhead{Ref.}& 
\colhead{Var.} & \colhead{Pol.} & \colhead{$\alpha$}&\colhead{X-ray} \\}
\startdata
J040349.35+261051.8&M2& 1 &	Class III& 1, 2	  &N&N&F&Y\\
J040443.07+261856.3&K4& 3 &	Class I  & 1, 2,3 &N&N&P&Y\\
J041314.16+281910.3&M4&	4 &	Class III& 1, 2   &Y&N&N&Y\\
J041327.23+281624.4&M0&	4 &	Class III& 1, 2   &Y&Y&N&Y\\
J041354.72+281132.6&K0-M3&5&	Class I  & 1, 3   &Y&N&P&Y\\
J041412.93+281211.9&K3& 6 & WTTS/Class II&1, 2, 7 &Y&N&P&Y\\
J041426.41+280559.4&M2.5/2.5&4&	 Class II&  8     &Y&N&F&Y\\
J041430.55+280514.4&K7& 4 &	Class II &  9     &N&N&P&Y\\
J041448.00+275234.4&M1&6, 10&	\nodata &\nodata&Y&N&P&Y\\
J041628.11+280735.4&K7V&11 &	\nodata &\nodata&Y&Y&N&N\\
J041829.10+282618.8&M1& 4 &	\nodata &\nodata  &N&N&N&Y\\
J041831.12+282715.9&K4& 12 &	Class III&1, 2    &N&Y&N&Y\\
J041833.38+283732.2&\nodata&\nodata&Class I&8&N&--&--&N\\
J041834.45+283029.9&M0& 4 &	Class II&8        &N&N&F&Y\\
J041840.62+281915.3&B8& 13 &	Class II&1, 2     &Y&Y&P&Y\\
J041847.04+282007.2&K7& 4 &	Class III&1, 2    &Y&Y&P&Y\\
J041851.49+282026.1&K7& 3 &	Class I  &1, 2, 3 &N&N&P&N\\
J041915.84+290626.6&K7& 14 &	Class II &1, 2    &N&N&P&Y\\
J041926.27+282614.0&K7& 4, 6&	Class III&1, 2    &Y&N&N&Y\\
J041941.28+274947.9&M0& 6, 10&  \nodata  &\nodata &Y&Y&P&Y\\
J042157.41+282635.3&K1& 1  &	Class II &1, 2    &N&N&P&Y\\
J042159.43+193205.7&K0& 1  &	Class II &1, 2, 3 &Y&Y&N&Y\\
J042200.70+265732.2&\nodata&\nodata&Class I&  15  &N&N&P&N\\
J042202.20+265730.3&M0-M3.5&3&	Class II & 3      &Y&N&N&Y\\
J042203.15+282538.8&M3& 1  &	Class III& 1, 2   &N&Y&F&Y\\
J042204.97+193448.3&M4.5&16& 	WTTS     & 16     &Y&N&F&Y\\
J042448.16+264315.9&K1 &16 &	WTTS	 & 16	  &N&N&F&Y\\
J042449.05+264310.2&K0 &16 &	Class III& 17     &Y&N&F&N\\
J042517.70+261750.2&K7 &15 & 	Class III& 15     &Y&N&P&Y\\
J042702.56+260530.4&\nodata&\nodata& Class II& 3  &N&N&P&N\\
J042704.69+260615.8&K6 & 4 &	Class II & 1, 3   &N&N&F&Y\\
J042920.70+263340.2&M4 & 4 & 	Class III&1, 2    &N&N&N&Y\\
J042923.74+243259.8&K3-7&1, 3 &	Class I  & 1, 2, 3&N&N&P&N\\
J042923.74+243301.3&K7-M2&18 &	\nodata&\nodata   &N&N&P&N\\
J042942.47+263249.0&M0& 4 &     Class III& 1      &Y&N&N&Y\\
J043114.45+271017.6&M0.5& 15&	Class III& 15     &Y&N&P&Y\\
J043138.42+181357.3&K5& 3 &	Class I&2, 3, 8   &N&--&P&Y\\
J043140.09+181356.7&M2/3.5& 19& Class II&1, 3     &Y&Y&N&Y\\
J043144.49+180831.6&\nodata&\nodata&Class I&1, 2, 3&N&N&P&Y\\
J043158.46+254329.8&M5.5& 4 &	Class III&1       &N&N&P&Y\\
J043214.58+182014.6&K7 & 4 &	Class III& 1, 2   &Y&Y&N&Y\\
J043215.42+242859.3&M0 & 3 & 	Class II & 3 	  &N&N&P&Y\\
J043215.86+180138.5&K7 & 4 & 	Class III& 1, 2   &N&N&F&Y\\
J043232.07+225726.3&K0-M4&5&	Class I&1, 3, 15  &Y&N&N&Y\\
J043242.82+255231.2&M2.0e+M3.0e&19&Class II&15    &N&N&P&Y\\
J043243.07+255230.7&M1V+M4V& 20&Class II&15       &Y&N&P&N\\
J043306.62+240954.8&K7/M3 & 15&  	Class II&15       &Y&N&F&Y\\
J043310.04+243343.1&K7 & 4 & 	Class III& 1, 2   &Y&N&P&Y\\
J043439.24+250100.9&K0 & 15&	Class III& 15     &Y&N&F&Y\\
J043520.92+225424.0&K7 &4, 15&	Class III&1, 2, 15&Y&N&P&Y\\
J043540.95+241108.6&M1 &4, 15&	Class II&1, 2, 15 &Y&N&P&Y\\
J043553.52+225408.9&K7 &4 &	Class III&1, 2    &Y&N&N&Y\\
J043935.20+254144.3&\nodata&\nodata&Class I&1, 2, 3&Y&N&P&Y\\
J043953.86+260309.7&\nodata&\nodata&Class 0&21    &N&N&F&N\\
J043955.75+254501.7&K4&3&	Class II&1, 2, 3  &N&N&P&Y\\
J044205.49+252256.0&K7&15&	Class III&15	  &Y&N&N&Y\\
J044207.32+252303.0&M1&15&	Class III&15      &N&N&N&Y\\
J045536.97+301754.8&K0&22&	WTTS&22           &Y&N&F&Y\\
J045545.85+303304.0&A0&23& 	Class II&1, 2     &N&N&P&Y\\
\enddata
\tablecomments{
Var.\ = Y when the source variability is higher than 50\% in at least one frequency; N when it is lower. Pol.\ = Y when circular polarization is detected and N when it is not. Sources in the outer fields for which the polarization could not be assessed are shown as "--". $\alpha$ refers to the spectral index, and is given as P (for positive) when it is higher than 0.2; F (for flat) when it is between --0.2 and $+$0.2, and N (for negative) when it is lower than --0.2. X-ray\ = Y when there is an X-ray flux reported in the literature, N when it is not.}
\tablerefs{
%
1 = Kenyon \& Hartmann~(1995); 2 = Andrews \& Williams~(2005); 3= White \& Hillenbrand~(2004); 4 = Brice\~no et al.~(2002);
5 = Connelley \& Greene (2010); 6 = Riviere-Marichalar et al.\ (2012); 7 = Hartigan et al.~(1994); 8 = Gutermuth et al.~(2009);
9 = Hartmann~(2002); 10 = Voges et al.~(1999); 11 = Devor et al.~(2008); 12 = White \& Ghez~(2001); 13 = Manoj et al.~(2006);
14 = Donati et al.~(2008); 15 = Rebull et al.~(2010); 16 = Mart\'in \& Magazz{\`u}~(1999); 17 = Wahhaj et al.~(2010);
18 = Doppmann et al.~(2008); 19 = Hartigan \& Kenyon (2003); 20 = Prato et al.~(2002); 21 = Chini et al.~(2001);
22 = Furlan et al.~(2011) and 23 = Hern\'andez et al.~(2004).
}
\end{deluxetable}

\newpage

\clearpage

\begin{center}
 {  Online Tables} 
\end{center}

\addtocounter{table}{-6} 
\begin{deluxetable}{ccrcrc}
\tabletypesize{\scriptsize}
\tablewidth{0pt}
\tablecolumns{6}
\tablecaption{Radio Sources Detected in Taurus-Auriga \label{tab:sources}}
\tablehead{           & \multicolumn{4}{c}{Flux Properties}	&Spectral\\
\colhead{GBS-VLA Name} & \colhead{$f_{4.5}$(mJy)\tablenotemark{a}} &\colhead{Var.\,(\%)} & \colhead{$f_{7.5}$(mJy)\tablenotemark{a}} & \colhead{Var.\,(\%)}&\colhead{Index}\\
}\startdata
J040331.11+260911.4&(1.45$\pm$0.16$\pm$0.07)$\times10^{+0}$&19.8$\pm$15.0&\nodata&\nodata&\nodata\\
J040345.47+261612.6&(5.54$\pm$1.17$\pm$0.28)$\times10^{-1}$&64.1$\pm$13.6&\nodata&\nodata&\nodata\\
J040349.35+261051.8&(3.45$\pm$0.23$\pm$0.17)$\times10^{-1}$&33.9$\pm$19.4&(3.62$\pm$0.24$\pm$0.18)$\times10^{-1}$&23.1$\pm$11.6&0.10$\pm$0.24\\
J040407.97+261402.1&(7.72$\pm$1.03$\pm$0.39)$\times10^{+0}$&\nodata&\nodata&\nodata&Extended\\
J040416.39+261544.5&(2.02$\pm$0.33$\pm$0.10)$\times10^{+0}$&36.2$\pm$17.6&\nodata&\nodata&\nodata\\
J040434.84+261810.7&(2.36$\pm$0.34$\pm$0.12)$\times10^{-1}$&35.7$\pm$21.8&(1.62$\pm$0.46$\pm$0.08)$\times10^{-1}$&$>$29.4$\pm$27.5&-0.76$\pm$0.66\\
J040443.07+261856.3&(2.06$\pm$0.20$\pm$0.10)$\times10^{-1}$&17.1$\pm$17.6&(3.31$\pm$0.19$\pm$0.17)$\times10^{-1}$&19.6$\pm$11.6&0.96$\pm$0.27\\
J040505.66+262311.5&(1.80$\pm$0.35$\pm$0.09)$\times10^{+0}$&44.2$\pm$16.2&\nodata&\nodata&\nodata\\
\enddata
\tablenotetext{a}{ The two given errors correspond to (i) statistical noise in the images and (ii) systematic uncertainty from possible errors in flux calibration.}
\tablenotetext{b}{Source not detected at three times the noise level on individual epochs,but detected on the image of the concatenated epochs.}
\tablenotetext{c}{Source detected only in the third epoch of Block A. The fluxes and spectral index 
are as detected in this epoch.}
\end{deluxetable}

\addtocounter{table}{+1} 
\begin{deluxetable}{lccccccc}
\tabletypesize{\scriptsize}
\tablewidth{0pt}
\tablecolumns{8}
\tablecaption{Radio Sources with known counterparts \label{tab:counterparts}}
\tablehead{           & Other         &         & \multicolumn{3}{c}{Infrared\tablenotemark{b}} &       & Object\\
\colhead{GBS-VLA Name} & \colhead{Names} &\colhead{X-ray\tablenotemark{a}} & \colhead{SST} & \colhead{2M}&\colhead{WISE}&
\colhead{Radio\tablenotemark{c}}&\colhead{type}\\}
\startdata
J040349.35+261051.8&2MASS J04034930+2610520&XEST&Y&Y&Y&--&YSO\\
J040416.39+261544.5&WISE J040416.33+261545.2&--&--&--&Y&--&--\\
J040434.84+261810.7&WISE J040434.83+261810.5&--&--&--&Y&--&--\\
J040443.07+261856.3&IRAS 04016+2610&XEST&--&Y&Y&--&YSO\\
J040505.66+262311.5&WISE J040505.68+262311.2&--&--&--&Y&NVSS&--\\
J041305.34+282255.1&WISE J041305.31+282255.1&--&Y&Y&Y&--&--\\
J041305.40+281413.9&--&--&--&--&--&NVSS&--\\
J041305.45+281414.7&--&--&--&--&--&NVSS&--\\
\enddata
\tablenotetext{a}{XEST = G\"udel et al. (2007); 1RXS = Voges et al. (1999); RX = Neuhaeuser et al.\,(1995) and 
BFR2003 = Bally et al.\,(2003).}
\tablenotetext{b}{SST = Padgett et al.\ (2007); 2M = Cutri et al.\ (2003) and WISE = Cutri et al.\ (2013).}
\tablenotetext{c}{
NVSS = Condon et al.\ (1998); BW = Becker \& White (1985); SBS = Skinner et al.\,(1993); OFMM~=~O'Neal~et~al.~(1990);
LRDRG = Loinard et al.\,(2007b); RAR = Rodr\'iguez et al.\,(1995); CB = Cohen \& Bieging (1986); 
ARC92 = Anglada et al.\,(1992); RRAB = Reipurth et al.\,(2004); B96 = Bontemps (1996); GRL2000 = Giovanardi et al.\,(2000)
and RR98 = Rodr\'iguez \& Reipurth\,(1998).
}
\end{deluxetable}


\begin{thebibliography}{}

\bibitem[Adelman-McCarthy 
\& et al.(2011)]{2011yCat.2306....0A} Adelman-McCarthy, J.~K., \& et al.\ 2011, VizieR Online Data Catalog, 2306, 0 

\bibitem[Andre et al.(1991)]{1991ApJ...376..630A} Andre, P., Phillips, 
R.~B., Lestrade, J.-F., \& Klein, K.-L.\ 1991, \apj, 376, 630 

\bibitem[Andrews 
\& Williams(2005)]{2005ApJ...631.1134A} Andrews, S.~M., \& Williams, J.~P.\ 2005, \apj, 631, 1134 

\bibitem[Anglada et al.(1992)]{1992ApJ...395..494A} Anglada, G., Rodriguez, 
L.~F., Canto, J., Estalella, R., \& Torrelles, J.~M.\ 1992, \apj, 395, 494

\bibitem[Anglada et al.(1998)]{1998AJ....116.2953A} Anglada, G., 
Villuendas, E., Estalella, R., et al.\ 1998, \aj, 116, 2953 

\bibitem[Bally et al.(2003)]{2003ApJ...584..843B} Bally, J., Feigelson, E., 
\& Reipurth, B.\ 2003, \apj, 584, 843

\bibitem[Becker 
\& White(1985)]{1985ApJ...297..649B} Becker, R.~H., \& White, R.~L.\ 1985, \apj, 297, 649

\bibitem[Benz 
\& Guedel(1994)]{1994A&A...285..621B} Benz, A.~O., \& Guedel, M.\ 1994, \aap, 285, 621

\bibitem[Briceno et al.(1993)]{1993PASP..105..686B} Briceno, C., Calvet, 
N., Gomez, M., et al.\ 1993, \pasp, 105, 686 

\bibitem[Brice{\~n}o et al.(1999)]{1999AJ....118.1354B} Brice{\~n}o, C., 
Calvet, N., Kenyon, S., \& Hartmann, L.\ 1999, \aj, 118, 1354 

\bibitem[Brice{\~n}o et al.(2002)]{2002ApJ...580..317B} Brice{\~n}o, C., 
Luhman, K.~L., Hartmann, L., Stauffer, J.~R., 
\& Kirkpatrick, J.~D.\ 2002, \apj, 580, 317 


\bibitem[Bontemps(1996)]{1996PhDT........72B} Bontemps, S.\ 1996, 
Ph.D.~Thesis, 

\bibitem[Carroll et 
al.(2012)]{2012A&A...548A..95C} Carroll, T.~A., Strassmeier, K.~G., Rice, J.~B., K{\"u}nstler, A.\ 2012, \aap, 548, A95


\bibitem[Chini et 
al.(2001)]{2001A&A...369..155C} Chini, R., Ward-Thompson, D., Kirk, J.~M., et al.\ 2001, \aap, 369, 155

\bibitem[Cohen 
\& Bieging(1986)]{1986AJ.....92.1396C} Cohen, M., \& Bieging, J.~H.\ 1986, \aj, 92, 1396 

\bibitem[Connelley 
\& Greene(2010)]{2010AJ....140.1214C} Connelley, M.~S., \& Greene, T.~P.\ 2010, \aj, 140, 1214

\bibitem[Condon et al.(1998)]{1998AJ....115.1693C} Condon, J.~J., Cotton, 
W.~D., Greisen, E.~W., et al.\ 1998, \aj, 115, 1693 

\bibitem[Cutri et al.(2003)]{2003tmc..book.....C} Cutri, R.~M., Skrutskie, 
M.~F., van Dyk, S., et al.\ 2003, ''The IRSA 2MASS All-Sky Point Source 
Catalog, NASA/IPAC Infrared Science 
Archive.~href=''http://irsa.ipac.caltech.edu/\\applications/Gator/''

\bibitem[Cutri 
\& et al.(2013)]{2013yCat.2328....0C} Cutri, R.~M., \& et al.\ 2013, VizieR Online Data Catalog, 2328, 0

\bibitem[Dallacasa et 
al.(2000)]{2000A&A...363..887D} Dallacasa, D., Stanghellini, C., Centonza, M., \& Fanti, R.\ 2000, \aap, 363, 887 

\bibitem[Damiani et al.(1994)]{1994ApJ...436..807D} Damiani, F., Micela, 
G., Sciortino, S., \& Harnden, F.~R., Jr.\ 1994, \apj, 436, 807 

\bibitem[Deller et 
al.(2013)]{2013A&A...552A..51D} Deller, A.~T., Forbrich, J., \& Loinard, L.\ 2013, \aap, 552, A51 

\bibitem[Devor et al.(2008)]{2008AJ....135..850D} Devor, J., Charbonneau, 
D., O'Donovan, F.~T., Mandushev, G., \& Torres, G.\ 2008, \aj, 135, 850

\bibitem[Dobashi et al.(2005)]{2005PASJ...57S...1D} Dobashi, K., Uehara, 
H., Kandori, R., et al.\ 2005, \pasj, 57, 1 

\bibitem[Donati et al.(2008)]{2008MNRAS.386.1234D} Donati, J.-F., Jardine, 
M.~M., Gregory, S.~G., et al.\ 2008, \mnras, 386, 1234 

\bibitem[Doppmann et al.(2008)]{2008ApJ...685..298D} Doppmann, G.~W., 
Najita, J.~R., \& Carr, J.~S.\ 2008, \apj, 685, 298 

\bibitem[Duch{\^e}ne et 
al.(2004)]{2004A&A...427..651D} Duch{\^e}ne, G., Bouvier, J., Bontemps, S., Andr{\'e}, P., \& Motte, F.\ 2004, \aap, 427, 651

\bibitem[Dzib et al.(2010)]{2010ApJ...718..610D} Dzib, S., Loinard, L., 
Mioduszewski, A.~J., Boden, A.~F., Rodr{\'{\i}}guez, L.~F., 
\& Torres, R.~M.\ 2010, \apj, 718, 610 

\bibitem[Dzib et al.(2011)]{2011ApJ...733...71D} Dzib, S., Loinard, L., 
Rodr{\'{\i}}guez, L.~F., Mioduszewski, A.~J., 
\& Torres, R.~M.\ 2011, \apj, 733, 71 

\bibitem[Dzib et al.(2013a)]{2013ApJ...775...63D} Dzib, S.~A., Loinard, L., 
Mioduszewski, A.~J., et al.\ 2013a, \apj, 775, 63

\bibitem[Dzib et al.(2013b)]{2013RMxAA..49..345D} Dzib, S.~A., 
Rodr{\'{\i}}guez, L.~F., Araudo, A.~T., 
\& Loinard, L.\ 2013b, \rmxaa, 49, 345 

\bibitem[Dzib et al.(2014)]{2014ApJ...788..162D} Dzib, S.~A., Loinard, L., 
Rodr{\'{\i}}guez, L.~F., \& Galli, P.\ 2014, \apj, 788, 162 

\bibitem[Elias(1978)]{1978ApJ...224..857E} Elias, J.~H.\ 1978, \apj, 224, 
857

\bibitem[Feigelson 
\& Montmerle(1999)]{1999ARA&A..37..363F} Feigelson, E.~D., \& Montmerle, T.\ 1999, \araa, 37, 363 

\bibitem[Feigelson et al.(1998)]{1998ApJ...494L.215F} Feigelson, E.~D., 
Carkner, L., \& Wilking, B.~A.\ 1998, \apjl, 494, L215 

\bibitem[Forbrich et 
al.(2007)]{2007A&A...469..985F} Forbrich, J., Massi, M., Ros, E., Brunthaler, A., \& Menten, K.~M.\ 2007, \aap, 469, 985 

\bibitem[Forbrich et al.(2010)]{2010ApJ...719..691F} Forbrich, J., Posselt, 
B., Covey, K.~R., \& Lada, C.~J.\ 2010, \apj, 719, 691 

\bibitem[Furlan et al.(2011)]{2011ApJS..195....3F} Furlan, E., Luhman, 
K.~L., Espaillat, C., et al.\ 2011, \apjs, 195, 3 

\bibitem[Gagn{\'e} et al.(2004)]{2004ApJ...613..393G} Gagn{\'e}, M., 
Skinner, S.~L., \& Daniel, K.~J.\ 2004, \apj, 613, 393 

\bibitem[Giardino et 
al.(2004)]{2004A&A...413..669G} Giardino, G., Favata, F., Micela, G., \& Reale, F.\ 2004, \aap, 413, 669

\bibitem[Giovanardi et al.(2000)]{2000ApJ...538..728G} Giovanardi, C., 
Rodr{\'{\i}}guez, L.~F., Lizano, S., \& Cant{\'o}, J.\ 2000, \apj, 538, 728 

\bibitem[G{\"u}del et 
al.(2007)]{2007A&A...468..353G} G{\"u}del, M., Briggs, K.~R., Arzner, K., et al.\ 2007, \aap, 468, 353

\bibitem[Guedel 
\& Benz(1993)]{1993ApJ...405L..63G} Guedel, M., \& Benz, A.~O.\ 1993, \apjl, 405, L63 

\bibitem[G{\"u}nther 
\& Schmitt(2009)]{2009A&A...494.1041G} G{\"u}nther, H.~M., \& Schmitt, J.~H.~M.~M.\ 2009, \aap, 494, 1041 

\bibitem[Gutermuth et al.(2009)]{2009ApJS..184...18G} Gutermuth, R.~A., 
Megeath, S.~T., Myers, P.~C., et al.\ 2009, \apjs, 184, 18 

\bibitem[Hartigan et al.(1994)]{1994ApJ...427..961H} Hartigan, P., Strom, 
K.~M., \& Strom, S.~E.\ 1994, \apj, 427, 961

\bibitem[Hartigan 
\& Kenyon(2003)]{2003ApJ...583..334H} Hartigan, P., \& Kenyon, S.~J.\ 2003, \apj, 583, 334 

\bibitem[Hartmann(2002)]{2002ApJ...578..914H} Hartmann, L.\ 2002, \apj, 
578, 914 

\bibitem[Heeschen(1984)]{1984AJ.....89.1111H} Heeschen, D.~S.\ 1984, \aj, 
89, 1111 

\bibitem[Heeschen 
\& Rickett(1987)]{1987AJ.....93..589H} Heeschen, D.~S., \& Rickett, B.~J.\ 1987, \aj, 93, 589

\bibitem[Hern{\'a}ndez et al.(2004)]{2004AJ....127.1682H} Hern{\'a}ndez, 
J., Calvet, N., Brice{\~n}o, C., Hartmann, L., 
\& Berlind, P.\ 2004, \aj, 127, 1682 

\bibitem[Kenyon 
\& Hartmann(1995)]{1995ApJS..101..117K} Kenyon, S.~J., \& Hartmann, L.\ 1995, \apjs, 101, 117 

\bibitem[Kenyon et al.(2008)]{2008hsf1.book..405K} Kenyon, S.~J., 
G{\'o}mez, M., 
\& Whitney, B.~A.\ 2008, Handbook of Star Forming Regions, Volume I, 405 

\bibitem[Koay et 
al.(2011)]{2011A&A...534L...1K} Koay, J.~Y., Bignall, H.~E., Macquart, J.-P., et al.\ 2011, \aap, 534, L1 

\bibitem[Kounkel et al.(2014)]{2014ApJ...790...49K} Kounkel, M., Hartmann, 
L., Loinard, L., et al.\ 2014, \apj, 790, 49 

\bibitem[Leinert et 
al.(1997)]{1997A&A...318..472L} Leinert, C., Richichi, A., \& Haas, M.\ 1997, \aap, 318, 472 

\bibitem[Loinard et al.(2005)]{2005ApJ...619L.179L} Loinard, L., 
Mioduszewski, A.~J., Rodr{\'{\i}}guez, L.~F., et al.\ 2005, \apjl, 619, 
L179

\bibitem[Loinard et al.(2007a)]{2007ApJ...671..546L} Loinard, L., Torres, 
R.~M., Mioduszewski, A.~J., et al.\ 2007a, \apj, 671, 546

\bibitem[Loinard et al.(2007b)]{2007ApJ...657..916L} Loinard, L., 
Rodr{\'{\i}}guez, L.~F., D'Alessio, P., Rodr{\'{\i}}guez, M.~I., 
\& Gonz{\'a}lez, R.~F.\ 2007b, \apj, 657, 916 

\bibitem[Loinard et al.(2008)]{2008ApJ...675L..29L} Loinard, L., Torres, 
R.~M., Mioduszewski, A.~J., 
\& Rodr{\'{\i}}guez, L.~F.\ 2008, \apjl, 675, L29 

\bibitem[Lovell et al.(2008)]{2008ApJ...689..108L} Lovell, J.~E.~J., 
Rickett, B.~J., Macquart, J.-P., et al.\ 2008, \apj, 689, 108 

\bibitem[Manoj et al.(2006)]{2006ApJ...653..657M} Manoj, P., Bhatt, H.~C., 
Maheswar, G., \& Muneer, S.\ 2006, \apj, 653, 657

\bibitem[Mart{\'{\i}}n 
\& Magazz{\`u}(1999)]{1999A&A...342..173M} Mart{\'{\i}}n, E.~L., \& Magazz{\`u}, A.\ 1999, \aap, 342, 173


\bibitem[Neuhaeuser et 
al.(1995)]{1995A&A...297..391N} Neuhaeuser, R., Sterzik, M.~F., Schmitt, J.~H.~M.~M., Wichmann, R., \& Krautter, J.\ 1995, \aap, 297, 391 

\bibitem[O'Dea(1998)]{1998PASP..110..493O} O'Dea, C.~P.\ 1998, \pasp, 110, 
493 

\bibitem[O'Neal et al.(1990)]{1990AJ....100.1610O} O'Neal, D., Feigelson, 
E.~D., Mathieu, R.~D., \& Myers, P.~C.\ 1990, \aj, 100, 1610 

\bibitem[Orienti
\& Dallacasa(2014)]{2014MNRAS.438..463O} Orienti, M., \& Dallacasa, D.\ 2014, \mnras, 438, 463 

\bibitem[Padgett et al.(2007)]{2007AAS...211.2904P} Padgett, D., McCabe, 
C., Rebull, L., et al.\ 2007, Bulletin of the American Astronomical 
Society, 39, 780 

\bibitem[Perley 
\& Butler(2013)]{2013ApJS..204...19P} Perley, R.~A., \& Butler, B.~J.\ 2013, \apjs, 204, 19 

\bibitem[Prato et al.(2002)]{2002ApJ...579L..99P} Prato, L., Simon, M., 
Mazeh, T., Zucker, S., \& McLean, I.~S.\ 2002, \apjl, 579, L99

\bibitem[Rebull et al.(2010)]{2010ApJS..186..259R} Rebull, L.~M., Padgett, 
D.~L., McCabe, C.-E., et al.\ 2010, \apjs, 186, 259 

\bibitem[Reipurth et al.(2004)]{2004AJ....127.1736R} Reipurth, B., 
Rodr{\'{\i}}guez, L.~F., Anglada, G., \& Bally, J.\ 2004, \aj, 127, 1736 

\bibitem[Riviere-Marichalar et 
al.(2012)]{2012A&A...538L...3R} Riviere-Marichalar, P., M{\'e}nard, F., Thi, W.~F., et al.\ 2012, \aap, 538, L3

\bibitem[Rodriguez et al.(1995)]{1995ApJ...454L.149R} Rodriguez, L.~F., 
Anglada, G., \& Raga, A.\ 1995, \apjl, 454, L149

\bibitem[Rodr{\'{\i}}guez 
\& Reipurth(1998)]{1998RMxAA..34...13R} Rodr{\'{\i}}guez, L.~F., \& Reipurth, B.\ 1998, \rmxaa, 34, 13 


\bibitem[Rodr{\'{\i}}guez et 
al.(2012a)]{2012A&A...537A.123R} Rodr{\'{\i}}guez, L.~F., Gonz{\'a}lez, R.~F., Raga, A.~C., et al.\ 2012a, \aap, 537, A123

\bibitem[Rodr{\'{\i}}guez et al.(2012b)]{2012RMxAA..48..243R} 
Rodr{\'{\i}}guez, L.~F., Dzib, S.~A., Loinard, L., et al.\ 2012b, \rmxaa, 
48, 243 

\bibitem[Rodr{\'{\i}}guez et al.(2014a)]{2014AJ....148...20R}
Rodr{\'{\i}}guez, L.~F., Carrasco-Gonz{\'a}lez, C., Montes, G.,
\& Tapia, M.\ 2014a, \aj, 148, 20

\bibitem[Rodr{\'{\i}}guez et al.(2014b)]{2014ApJ...793L..21R} 
Rodr{\'{\i}}guez, L.~F., Zapata, L.~A., Dzib, S.~A., et al.\ 2014b, \apjl, 
793, L21 


\bibitem[Skinner et al.(1993)]{1993ApJS...87..217S} Skinner, S.~L., Brown, 
A., \& Stewart, R.~T.\ 1993, \apjs, 87, 217

\bibitem[Smith et 
al.(2005)]{2005A&A...431..307S} Smith, K.~W., Balega, Y.~Y., Duschl, W.~J., et al.\ 2005, \aap, 431, 307 

\bibitem[Stelzer et al.(2005)]{2005ApJS..160..557S} Stelzer, B., Flaccomio, 
E., Montmerle, T., et al.\ 2005, \apjs, 160, 557 

\bibitem[Stelzer et 
al.(2006)]{2006A&A...457..223S} Stelzer, B., Micela, G., Hamaguchi, K., \& Schmitt, J.~H.~M.~M.\ 2006, \aap, 457, 223

\bibitem[Stelzer et 
al.(2009)]{2009A&A...493.1109S} Stelzer, B., Robrade, J., Schmitt, J.~H.~M.~M., \& Bouvier, J.\ 2009, \aap, 493, 1109 

\bibitem[Torres et al.(2007)]{2007ApJ...671.1813T} Torres, R.~M., Loinard, 
L., Mioduszewski, A.~J., \& Rodr{\'{\i}}guez, L.~F.\ 2007, \apj, 671, 1813

\bibitem[Torres et al.(2009)]{2009ApJ...698..242T} Torres, R.~M., Loinard, 
L., Mioduszewski, A.~J., \& Rodr{\'{\i}}guez, L.~F.\ 2009, \apj, 698, 242

\bibitem[Torres et al.(2012)]{2012ApJ...747...18T} Torres, R.~M., Loinard, 
L., Mioduszewski, A.~J., et al.\ 2012, \apj, 747, 18 

\bibitem[Voges et 
al.(1999)]{1999A&A...349..389V} Voges, W., Aschenbach, B., Boller, T., et al.\ 1999, \aap, 349, 389

\bibitem[Wahhaj et al.(2010)]{2010ApJ...724..835W} Wahhaj, Z., Cieza, L., 
Koerner, D.~W., et al.\ 2010, \apj, 724, 835

\bibitem[White 
\& Hillenbrand(2004)]{2004ApJ...616..998W} White, R.~J., \& Hillenbrand, L.~A.\ 2004, \apj, 616, 998 

\bibitem[White 
\& Ghez(2001)]{2001ApJ...556..265W} White, R.~J., \& Ghez, A.~M.\ 2001, \apj, 556, 265 


\bibitem[Zinnecker 
\& Preibisch(1994)]{1994A&A...292..152Z} Zinnecker, H., \& Preibisch, T.\ 1994, \aap, 292, 152 


\end{thebibliography}
\end{document}